\documentclass[11pt]{article}
\usepackage{epsfig}
\usepackage{amsfonts}
\usepackage{mathrsfs}
\usepackage{latexsym}
\usepackage{color}
\usepackage[usenames,dvipsnames]{xcolor}
\usepackage{amsmath,amssymb}
\usepackage{verbatim}
\definecolor{red}{rgb}{1,0,0}
\usepackage{cite}
\numberwithin{equation}{section}
\def\bea{\begin{eqnarray}} 
\def\eea{\end{eqnarray}}
\def\be{\begin{equation}} 
\def\ee{\end{equation}} 
\def\ba{\begin{array}}
\def\ea{\end{array}} 
\def\nn{\nonumber}
\newcommand{\vs}[1]{\vspace{#1 mm}}

\begin{document}

\thispagestyle{empty}

\renewcommand{\thefootnote}{\fnsymbol{footnote}}
\setcounter{footnote}{0}
\setcounter{figure}{0}

\begin{titlepage}

\begin{center}

{\Large{\textbf{Covariant and single-field effective action with the background-field formalism}}}
\vs{10}

{\large 
Mahmoud Safari\footnote{e-mail address: safari@bo.infn.it} 
and Gian Paolo Vacca\footnote{e-mail address: vacca@bo.infn.it}
}\\
\vs{5}

Dipartimento di Fisica and INFN - Sezione di Bologna, 
via Irnerio 46, 40126 Bologna, Italy.

\vs{10}

\begin{abstract}

In the context of scalar quantum field theory we introduce a class of generically nonlinear quantum-background splits for which the splitting 
Ward identity, encoding the single field dependence in the effective action, can be solved exactly.
We show that this can be used to construct an effective action which is both covariant and dependent on the background and fluctuation fields only through a single total field in a way independent from the dynamics.
Moreover we discuss the criteria under which the ultraviolet symmetries are inherited by the quantum effective action.
The approach is demonstrated through some examples, including the $O(N)$ effective field theory, which might be of interest for the Higgs sector of the Standard Model or its extensions.

\end{abstract}

\end{center}





\end{titlepage}

\setcounter{page}{2}
\renewcommand{\thefootnote}{\arabic{footnote}}
\setcounter{footnote}{0}


\section{Introduction}

The effective action of quantum field theories, as the generator of 1PI vertices, provides the building blocks for constructing scattering amplitudes,
from which physical information such as cross sections are extracted. It is well known that physical observables, for example constructed from S-matrix elements, should be in general invariant under field reparameterization. 
In the recent years there has been a strong focus on developing and applying on-shell methods which are expecially useful for gauge theories.
On the other side we note that implications of global symmetries at quantum level  are generally investigated in an off-shell framework.
For example spontaneous symmetry breaking analysis is usually not done with S-matrix elements but using the off-shell effective action (or effective potential) as in the
Coleman-Weinberg approach~\cite{Coleman:1973jx}.
In a standard definition the quantum effective action is not a scalar under field reparametrizations but nevertheless the S-matrix elements, constructed with the well known Lehmann-Symanzik-Zimmermann procedure
involving the asymptotic on-shell fluctuations, are invariant under any well behaved reparametrization ($\phi'=\phi+f(\phi)$) which transforms one set of coordinates to another in the configuration field target manifold.
Indeed this is not the case for the off-shell n-point connected Green functions constructed from the proper vertices which change from one parametrization to another by terms
which vanish when evaluated on the vacuum configuration and acting on the external asymptotic fluctuations (particles).

The desire of having an action invariant under reparametrization had already appeared long ago in the context
of nonlinear sigma models~\cite{honerkamp_npb1972,honerkamp_npb1974,alvarez_annp1981,boulware_annp1982,mukhi},
since the target space field manifold has no really preferred chart for parametrizing the fields.
There is a well known covariant construction, whose formulation was completed by Vilkovisky and DeWitt~\cite{vilkovisky_gospel1984,vilkovisky_npb1984,dewitt_effact}, and studied further by other authors including \cite{FT,rebhan_npb1987,rebhan_npb1988,burgess_kunstatter,ET},
which allows to obtain an effective action which is a functional scalar under field reparametrization~\footnote{See also~\cite{BV83,BV85,BO,Od}.}.
Such a description, if one is interested in the computation of S-matrix elements, is clearly unnecessary even if more elegant.
On the other hand there are cases where a covariant off-shell description is certainly desirable.
Apart from the aforementioned analysis of the global symmetries and the vacuum,
it may be convenient  when dealing with an effective field theory approach or with exact Wilsonian renormalization methods.
Moreover covariant approaches in target space were also considered in the studies of the quantum symmetries in string theory models~\cite{Callan:1989nz}.

Such an approach is geometrical and based on the introduction of a connection on the configuration space manifold.
The QFT generating functionals are obtained by introducing a source coupled to a quantum fluctuation, transforming as a contravariant vector of the target manifold, 
which results in a highly nonlinear dependence on the field.
Because of the geodesic construction (typically from an exponential map),
in general the covariant effective action depends on a base point field (background field) and on the (average) quantum field fluctuation.

In general a nonlinear splitting of a quantum field into a background and a quantum fluctuation leads to a double-field dependence in the effective action,
as is typical for example in the covariant analysis of nonlinear $\sigma$-models, where here a ``field'' generically refers to a number of fields.
This fact is very inconvenient and questions the main advantage of the background-field method since from a background field computation, 
willing to keep the covariance, one cannot reconstruct the full quantum field dependence in the effective action functional.

In this work we therefore address this problem and determine the conditions and the procedures which one should use to construct a single-field dependent effective action, which is also naturally covariant.
In particular we shall consider a class of quantum-background splits for which the splitting Ward identity (spWI) simplifies significantly to the extent that it can be solved exactly.
It turns out that for such splittings any function of the total field will solve the spWI, with a total field that depends on the background and the fluctuation field in a local way, independent of the QFT dynamics, and in fact in the same way as that in the ultraviolet.
However, the splitting is still found to be general enough to allow for the construction of a covariant effective action. 

An important point which remains to be addressed regards the symmetry properties of the effective action. Indeed in the standard quantization procedure the quantum effective action may not inherit
the ultraviolet symmetries of the bare action~\footnote{We shall consider in the present work QFT models free from anomalies.}.
Here we shall discuss, making use of old results of Coleman, Wess and Zumino \cite{cwz}, that for a certain class of symmetries, including linear and nonlinear ones, the covariant single-field effective action
may enjoy the symmetry properties of the ultraviolet defining theory. 

More recently the covariant approach to non linear sigma models has also received attention in applications to the effective field theory
of the Higgs boson~\cite{manohar_1511,manohar_1602,manohar_1605}, which might be of interest for LHC phenomenology. 
While having the benefit of preserving the ultraviolet symmetries in the effective action, the applications of the covariant formalism suffer from double-field dependence, that is, an explicit dependence on the background field in addition to the average quantum field, and the results are mainly valid only at the background level. Our general results can therefore be of interest in such applications.

Another field of application of the techniques developed here regards renormalization group analysis.
One interesting framework consists in the functional renormalization  group (FRG) approach inspired by the work of Wilson.
We address this problem in a separate work \cite{SV}. The covariance and single-field properties will severely constrain the set of operators that would otherwise be present in the effective action. The Standard Vilkovisky-DeWitt approach in the FRG context has been discussed in \cite{JP03,JP05,Safari}.

In this work we concentrate on scalar theories and do not address the more subtle problem of gauge theories, that involves a gauge fixing procedure which introduces an explicit background field dependence in the off-shell effective action. 

Our work is organized as follows.
In Section 2 we setup the problem and review the Ward identity condition for the single-field dependence of the effective action. We then introduce a special class of quantum-background splits 
and solve the corresponding spWI in this case. The resulting effective action is shown to be covariant and manifestly background-independent. 
After dealing with the more concrete construction of the one loop effective action we then discuss at a general level the symmetries of the effective action. Specifically, we classify the ultraviolet symmetries that, in our construction, are preserved in the effective action.
In Section 3 we discuss in some details two examples: a single-scalar QFT and the $O(2)$ linear model in flat field space.
In Section 4  we address the case of a nonlinear $\sigma$-model giving all the details necessary to construct the one-loop covariant and single-field dependent effective action.
In Section 5 we analyze general $O(N)$ invariant $N$-scalar theories for which the target space is not necessarily flat, and in the Vilkovisky-DeWitt approach the background result is not enough to reconstruct the fluctuation dependence. 
Such theories may be of interest for the extensions of the Higgs sector of the Standard Model.
After the conclusion we have also added two appendices. In the first we show how the splitting in background and quantum fields we have introduced is related to the exponential splitting, 
and in the second, for a general nonlinear sigma model, we have given a more direct check of the single-field dependence of the one-loop effective potential.

\section{Covariant and single-field effective action} 
In this section we shall consider a bosonic (non-gauge) quantum field theory and search for possibly non linear splittings of the quantum field
into a background and a quantum fluctuation which, by means of a usual path integral quantization method,
leads to a quantum effective action which is manifestly background independent, i.e. can be written as a functional of a single total field. This will also serve as a basis for a subsequent work in the context of FRG \cite{SV}.
Our goal is to obtain a description which is also covariant, that is to find an off-shell effective action which transforms as a scalar under field reparameterizations.
Before getting into the main discussion let us review briefly the notion of 
spWI. 

\subsection{Splitting Ward identities}
Let us consider the quantization of a bosonic theory with bare action $S[\phi]$, when the field (multiplet) $\phi^i=\phi^i(\varphi,\xi)$ is split into a background field $\varphi^i$ and a quantum field $\xi^i$.
The generator of connected $n$-point functions $W[\varphi,J]$
is a functional of the background and a source field $J_i$ coupled to the quantum field $\xi^i$, and is given in Euclidean space by the path integral 
\be  \label{W}
e^{-W[\varphi,J]} = \int \!\! D\phi \;\mu(\phi) \,\; e^{-S[\phi]-J\cdot\xi}.
\ee
We have allowed for a path integral measure that depends only on the field $\phi^i$ and not separately on the quantum field $\xi^i$.
Deviations from total field dependence in the measure will be irrelevant for example when using dimensional regularization. 
As usual, on performing a Legendre transform, one defines the generator of the 1PI vertices,
the effective action:
\be \label{eaa_def}
\Gamma[\varphi,\bar\xi] = W[\varphi,J] - J\!\cdot\!\bar\xi,
\ee
with $\bar \xi=\langle \xi \rangle$, which in general has a dependence on both the background and the fluctuation field. The effective action satisfies the following functional integro-differential equation:
\be
\label{ea_equation}
e^{-\Gamma[\varphi,\bar\xi]} = \int \!\! D\phi \;\mu(\phi) \, e^{-S[\phi]+\Gamma_{; i} (\xi-\bar\xi)^i }.
\ee
Here a semicolon ``;'' denotes a derivative with respect to the quantum field $\xi$ while a comma
will be used for the derivative with respect to the background field $\varphi$ and in general whenever convenient we shall use the DeWitt condensed notation.

Taking a functional derivative of this equation with respect to the background field $\varphi^i$
one obtains the splitting Ward identity~\cite{rebhan_npb1988,hps,Blasi:1988sh,burgess_kunstatter,kunstatter_1992,parker_toms,Safari}:
\be
\label{sWI}
\Gamma_{,i}+ \Gamma_{;j} \langle \xi^j_{,i}\rangle =0.
\ee

In the next subsection we will discuss some possible quantum-background splittings for which the above equation can be easily solved.
In such cases where the Ward identity is solved the effective action can be explicitly written in terms of a total field $\bar \phi(\varphi,\bar \xi)$, with no extra background dependence.

\subsection{Flat quantum-background split} \label{fqbs}
The solution to the splitting Ward identity for the effective action $\Gamma$ given in Eq.~(\ref{sWI}) can be in general extremely involved with a non trivial dependence on the particular dynamics of the model considered. 
The source of complication is the average quantity $\langle \xi^j_{,i}\rangle$ appearing in the equation.
In general this term is a highly nonlocal function of the quantum and background fields, with an implicit dependence on the effective action itself.
This makes solving the equation very difficult if not impossible. In order to avoid this complicated dependence of $\langle \xi^j_{,i}\rangle$
on the fields and the effective action, it is sufficient to require $\xi^j_{,i}$ to depend at most linearly on the fluctuation fields.
This way we avoid two-point and higher correlation functions to appear in the expression for $\langle \xi^j_{,i}\rangle$. 
With this requirement the most general form $\xi^j_{,i}$ can take is
\be \label{linear split}
\xi^k\!\!,_i = \alpha^k_i(\varphi) - \beta^k_{ij}(\varphi)\,\xi^j.
\ee
In this case the average quantity simplifies $\langle\xi^k\!\!,_i\rangle = \bar\xi^k\!\!,_i$, and the splitting Ward identity reduces to
\be \label{glspwi} 
\Gamma,_i + \Gamma_{;k}\,\bar\xi^k\!\!,_i=0
\ee
which admits the general solution $\Gamma[\phi(\varphi,\bar\xi)]$, where $\xi(\varphi,\phi)$ is a solution to \eqref{linear split}. 
This is because the first order differential operator acting on $\Gamma$ in \eqref{glspwi} is simply the partial background derivative keeping the total field fixed,
and therefore the solutions to this equation consist of functionals of the total field $\bar\phi=\phi(\varphi,\bar\xi)$.
Equation \eqref{linear split} is solvable if and only if, $\alpha^k_i$ and $\beta^k_{ij}$, regarded as tensor valued one-forms, satisfy
\be  
\label{int_cond}
d\alpha^k + \beta^k_j\wedge\alpha^j=0, \qquad d\beta^k_j + \beta^k_l\wedge\beta^l_j=0.
\ee
These are simply the Frobenius conditions for Eq.~\eqref{linear split} $\xi^k\!\!,_{[ij]}=0$, where as in the equation itself, the derivatives are taken keeping the total field fixed.
Of course, one would obtain the same conditions \eqref{int_cond} by plugging the right hand side of \eqref{linear split}
in \eqref{glspwi} and imposing the condition that the commutator of the differential operator
in \eqref{glspwi} vanish. The solution to the equations (\ref{int_cond}), which are the same as the zero torsion and curvature Cartan structure equations, is
\be 
\beta^k_{ij} = (U^{-1})^k_a\partial_i U^a_j, \qquad \alpha^k_i = -(U^{-1})^k_a\partial_i f^a,
\ee
where $f$ is a vector-valued function and $U$ is a matrix-valued function of the background field. 
The minus sign in the solution for $\alpha$ is there for convenience. With the integrability conditions \eqref{int_cond}, Eq.~\eqref{linear split} can be solved to give
\be 
\xi^k(\varphi,\phi) = -(U^{-1}(\varphi))^k_a\, \left(f^a(\varphi) - (g^{-1})^a(\phi)\right), \qquad \mathrm{or} \qquad \phi^k(\varphi,\xi) = g^k\left[f(\varphi) + U(\varphi)\xi\right]
\label{split1}
\ee
where $g$ is an arbitrary function. 
We have therefore shown that the effective action $\Gamma$ solution of Eq.~(\ref{glspwi}) can be considered as a function of the single field $\bar \phi^k(\varphi,\bar\xi)$, which is a function of the background and quantum fields in the special form given above. 
This functional dependence of $\bar \phi^k$ on $\varphi^i,\bar\xi^i$ is the same as that in the ultraviolet theory and is unaltered through the functional quantization procedure.
From the above relations, it is also clear that the choice $g=f^{-1}$ corresponds to the boundary condition $\xi^k(\varphi,\varphi)=0$ or $\phi^k(\varphi,0)=\varphi^k$.
In such a case the quantum-background split introduced here is in fact related to the exponential map $[Exp_\varphi \xi]_\Gamma$, defined with a flat connection $\Gamma^k_{ij}$. 
This is discussed explicitly in Appendix A, where we show that the construction is associated to the flat connection\footnote{For quantities depending on a single field, a comma should be understood as the derivative with respect to their argument.}
\be
 \Gamma^k_{ij}= (f^{-1})^k\!\!,_b f^b\!\!,_{ij}  = -(f^{-1}),_{ab}^k\,f^a\!\!\!,_i\,f^b\!\!\!,_j
 \label{flatconn}
\ee
and therefore give it the name ``flat splitting''. 

\subsection{Covariance}

We want to stress that this ``exponential'' splitting can be used to define an effective action covariant with respect to field reparametrization.
This can be seen without referring to the exponential expansion. Under a coordinate transformation $h$ the total field is expected to transform as $\phi\rightarrow h(\phi)$
{\setlength\arraycolsep{2pt}
\bea
(f^{-1})^i\left(f(\varphi)+ U\xi\right) &\rightarrow & h((f^{-1})^i\left(f(\varphi)+ U\xi\right)) \nn\\
&=& ((f\!\circ\! h^{-1})^{-1})^i\left((f\!\circ\! h^{-1})(h(\varphi))+ U(\partial h)^{-1}\partial h\xi\right). 
\eea}%
This means that $\phi\rightarrow h(\phi)$ follows from
\be 
\varphi\rightarrow h(\varphi),\quad
\xi\rightarrow \partial h\,\xi,\quad
f\rightarrow f\!\circ\! h^{-1},\quad
U\rightarrow U(\partial h)^{-1}.
\label{cov_split}
\ee
In particular, the quantum field $\xi$ is seen to transform linearly under a change of coordinates.
This implies that the effective action is invariant under the above transformations, i.e. it is covariant.
According to the single-field property of the effective action the background and fluctuation dependencies are collected into its dependence on the total field $\phi$.
However there can in principle be a separate dependence on the function $f$, or equivalently the flat connection we have introduced.
This is explicitly seen in the one-loop effective action discussed in the next subsection. Also, a $U$ dependence can appear only implicitly through $\phi$.
The reason is that the ultraviolet action has a symmetry $\xi\rightarrow A\xi$, $U\rightarrow U A^{-1}$, where $A$ is any matrix valued function of the background.
This is also a symmetry of the effective action, so a supposed explicit $U$ dependence would be removed by such a transformation with $A=U$, and therefore it cannot appear explicitly.
For this reason any choice for the matrix $U$ which has the right transformation property given in \eqref{cov_split} will do the job.
In particular $U=\partial f$ is a natural choice in the sense that it is related to the already existing function $f$, and transforms in the correct way under a field redefinition.
This choice also leads exactly to the exponential splitting with a flat connection as discussed in the appendix.
So, except for the next subsection where we would like to show explicitly how $U$ drops out in the final expression for the one-loop effective action,
for the rest of the paper we will stick to this choice, and by ``flat splitting'' we refer to 
\be 
\phi^k(\varphi,\xi) = (f^{-1})^k\left[f^a(\varphi) + f,_i^a(\varphi)\,\xi^i\right].
\label{fs} 
\ee
The covariance and single-field properties of the effective action are therefore summarized in the following equation
\be \label{cov}
\Gamma_{G',f'}[\phi'] = \Gamma_{G,f}[\phi], \qquad (\mathrm{or}\;\, \Gamma'[\phi'] = \Gamma[\phi], \quad \Gamma'\equiv \Gamma_{G',f'}, \; \Gamma\equiv \Gamma_{G,f})
\ee
where we have made explicit the dependence of the effective action on $G$ which represents not only the metric but all the field-space tensors present in the ultraviolet action $S_G[\phi]$, and also the dependence on the flat connection through the function $f$.

Let us stress here that in such a case the vector $\xi$, which satisfies also Eq.~(\ref{geodesiceq}), is generically written as
\be 
\xi^k(\varphi,\phi) = \left[(\partial f(\varphi))^{-1}\right]^k_a\, \left[f^a(\phi) -f^a(\varphi) \right]
\label{solgeoeq}
\ee
and transforms covariantly.
In particular it can be seen originating from a standard linear splitting (corresponding to the case $f={\rm id}$) followed by a reparameterization of the fields.
The covariant dependence on $f$, as well as the dependence in the total field only, can also be seen explicitely by rewriting Eq.~(\ref{ea_equation}) as
\be
\label{alt_ea_equation}
e^{-\Gamma[\bar\phi]} = \int \!\! D\phi \;\mu(\phi) \, e^{-S[\phi]+\frac{\delta\Gamma}{\delta \bar\phi^i} \left[(\partial f)^{-1}(f(\bar\phi))\right]^i_a \left[ f(\phi)-f(\bar \phi)\right]^a },
\ee
from which one can directly see that this expression can be obtained also performing a change of variable on the case of the standard linear splitting ($f={\rm id}$) according to Eq.~(\ref{cov_split}).
As we will show in the next Sections we shall be able to give a prescription for the choice of the function $f$ which allows for the UV symmetries to be preserved, making
the dependence of the effective action $\Gamma$ on $f$ not an issue.

\subsection{One-loop effective action} \label{olea}

We will demonstrate the rather abstract ideas of the previous sections through the explicit computation of the effective action at one-loop. The general expression for the one-loop effective action in the background-field formalism is
\be \label{1loopeffact}
\Gamma^{1-\mathrm{loop}} = S[\phi] + {\textstyle{\frac{i}{2}}}\mathrm{Tr}\log{ S^{(2)}[\phi]}
\ee
where $S^{(2)}$ is the second fluctuation derivative of the ultraviolet action. Before entering into the explicit computation of $S^{(2)}$ for flat splitting, let us consider a general split $\phi(\varphi,\xi)$. In this case the quantity $S^{(2)}$ can be written as
\be \label{s2general}
S_{;ij}[\phi(\varphi,\xi)]  = S,_{pq} \phi^p\!\!_{;i}\phi^q\!\!_{;j} + S,_p \phi^p\!\!_{;ij} .
\ee
Now let us consider an exponential splitting with an arbitrary connection whose first few terms in an expansion in $\xi$ are 
\be 
\phi^i=\varphi^i+\xi^i -{\textstyle{\frac{1}{2}}}\, \Gamma^{i}_{pq}(\varphi)\,\xi^p\xi^q + \cdots 
\ee
Using this expansion in \eqref{s2general} one can see the well known fact that at leading order, 
setting $\xi=0$ in a $\xi$ expansion, the second fluctuation derivative of the action is nothing but its second covariant derivative at the background  
\be 
S_{;ij}[\varphi] = \nabla_i\nabla_j S[\varphi].
\ee
One  may expect that for $\xi\neq 0$ this relation generalizes to
\be \label{1}
S_{;ij}[\phi] = \nabla_p\nabla_q S[\phi]\, \phi^p\!\!_{;i}\phi^q\!\!_{;j} .
\ee
This is in fact not the case, and it is important to notice that \eqref{1} does not necessarily continue to hold beyond leading order. 
Instead, as we will now see, for flat splitting this identity is valid at all orders in the fluctuation field, that is, at the level of the full (total) field. 
In this case from the general relation \eqref{s2general} and the explicit form of flat splitting 
\be \label{fsp}
\phi^i(\varphi,\xi) = (f^{-1})^i\left(f(\varphi)+ U\xi\right) 
\ee
and its first two derivatives 
\be 
\phi^p\!\!_{;i}=(f^{-1})^p_{,a}( f(\phi) ) U^a_i(\varphi), \qquad 
\phi^p\!\!_{;ij}=(f^{-1})^p_{,ab}( f(\phi) ) U^a_i(\varphi)U^b_j(\varphi)
\ee
we have
{\setlength\arraycolsep{2pt}
\bea
S_{;ij}[\phi(\varphi,\xi)] 
&=& S,_p (f^{-1})^p\!\!,_{mn}U^m_iU^n_j + S,_{pq} (f^{-1})^p\!\!,_m U^m_i (f^{-1})^q\!\!,_n U^n_j \nn\\
&=&
\big[S,_{pq}-\Gamma^k_{pq} S,_k\big] (f^{-1})^p\!\!,_m U^m_i (f^{-1})^q\!\!,_n U^n_j \nn\\
&=&
\nabla_p\nabla_q S\;\, (f^{-1})^p\!\!,_m U^m_i (f^{-1})^q\!\!,_n U^n_j, \label{second derivative}
\eea}%
where here the connection $\Gamma^k_{pq}$ is defined in Eq.~(\ref{flatconn}).
In the above, the argument of the derivatives of $(f^{-1})^p$ is $f(\phi)$, and the argument of $f$ is $\phi$. 
Recall that in this case a comma denotes differentiation with respect to the single argument $\phi$. 
We stress that $\Gamma^k_{ij}$ is a flat connection which vanishes after a change of coordinates by $f$: For a general transformation $\bar U$, a connection $C^k_{ij}$, considered also as a matrix-valued one-form $(C_i)^k_j$, transforms as
\be 
C^k_{ij} \rightarrow (\bar U^{-1})^a_i\, \bar U\left[C_a+\partial_a\right]\bar U^{-1} .
\ee
So a vanishing connection will be transformed to
\be 
0 \rightarrow (\bar U^{-1})^a_i\, \bar U\left[0+\partial_a\right]\bar U^{-1} = (\bar U^{-1})^a_i\, \bar U^k_b\,\partial_a (\bar U^{-1})^b_j = - (\bar U^{-1})^a_i\,(\bar U^{-1})^b_j\, \partial_a \bar U^k_b 
\ee
therefore, under the coordinate transformation $x^i\rightarrow x'^i = (f^{-1})^i(x)$, we have $\bar U^i_a = \partial x'^i/\partial x^a$ and the resulting connection will be
\be 
0 \rightarrow  - \frac{\partial x^a}{\partial x'^i}\,\frac{\partial x^b}{\partial x'^j}\, \frac{\partial x'^k}{\partial x^a\partial x^b} \equiv \Gamma^k_{ij}(x).
\ee
The factor on the r.h.s of \eqref{second derivative} is canceled by the Jacobian
\be 
\frac{\delta\phi^i}{\delta\xi^p} = (f^{-1})^i\!\!,_a U^a_p
\ee
and therefore the one-loop effective action is given by
\be \label{1loopea}
\Gamma^{1-\mathrm{loop}}_{\!f}[\phi] = S[\phi] + {\textstyle{\frac{i}{2}}}\mathrm{Tr}\log\left[S,_{ij}-\Gamma^k_{ij} S,_k\right].
\ee
Notice that for the case of an exponential expansion with a flat connection $U=\partial f(\varphi)$, the Jacobian and the extra term on the r.h.s of \eqref{second derivative} does not equal identity because $(f^{-1})^i\!\!,_a$ is evaluated at $f(\phi)$ while $U^a_p = f\!,_p^a(\varphi)$ is evaluated at $\varphi$. For $\xi =0$ we would have $\phi = \varphi$ and this would give an identity $(f^{-1})^i\!\!,_a(f(\varphi))f\!,_j^a(\varphi) = \delta^i_j$. This effective action depends on a single field $\phi$ through $S[\phi]$ and its derivatives and also the flat connection $\Gamma^k_{ij}[\phi]$. This is the benefit of using a parametrization of the form \eqref{split1}. Notice that this kind of splitting includes the $Exp$ splitting with a flat connection when $U=\partial f$. For other kind of splittings $\phi(\varphi,\xi)$, like the $Exp$ splitting with a non-flat connection, which also gives rise to a covariant effective action, the simple single-field dependence is lost. In such a case the Christoffel symbol in \eqref{1loopea} is replaced with  
\be 
- \frac{\partial \xi^a}{\partial \phi^i}\,\frac{\partial \xi^b}{\partial \phi^j}\, \frac{\partial \phi^k}{\partial \xi^a\partial \xi^b}
\ee
which does not depend on a single field $\phi$.

\subsection{Single-field dependence} \label{sfd}

The general discussions of Sections~\ref{fqbs} and~\ref{olea} show that the exponential splitting based on a flat connection leads to an effective action that manifestly depends on a single field.
It would still be instructive to see this, and in particular the role of the flat connection, explicitly at the one-loop level.
For this purpose let us consider the exponential expansion in the fluctuation fields, of the action based on a general connection $\nabla$
\be 
S[\phi] = S[\varphi]+\xi^p\nabla_p S[\varphi] + {\textstyle{\frac{1}{2!}}}\,\xi^p\xi^q\nabla_p\nabla_q S[\varphi] + {\textstyle{\frac{1}{3!}}}\,\xi^p\xi^q\xi^r\nabla_{(p}\nabla_q\nabla_{r)} S[\varphi] +
{\textstyle{\frac{1}{4!}}}\,\xi^p\xi^q\xi^r\xi^s\nabla_{(p}\nabla_q\nabla_r\nabla_{s)} S[\varphi] \cdots
\ee
From this, the second fluctuation derivative, at non zero fluctuating fields, is
\be 
S_{;ij}[\phi] = \nabla_i\nabla_j S[\varphi] + \xi^p\nabla_{(p}\nabla_i\nabla_{j)} S[\varphi] +{\textstyle{\frac{1}{2!}}}\,\xi^p\xi^q\nabla_{(p}\nabla_q\nabla_i\nabla_{j)} S[\varphi] + \cdots
\label{S2fluct}
\ee
This is what appears in the argument of the logarithm in the expression for the one-loop effective action \eqref{1loopeffact}. At the background level this simplifies to $\nabla_i\nabla_jS[\varphi]$.
We would like to see if one can reconstruct the one-loop effective action \eqref{1loopeffact} by taking the background result
\be \label{1loopeffactbg}
\Gamma^{1-\mathrm{loop}}[\varphi,\xi=0] = S[\varphi] + {\textstyle{\frac{i}{2}}}\mathrm{Tr}\log \nabla\nabla S[\varphi]
\ee
and promoting the background field to the total field. To find out, let us expand in the fluctuation field the second covariant derivative of the action evaluated at the total field
{\setlength\arraycolsep{2pt}
\bea
\nabla_i\nabla_j S[\phi] &\stackrel{*}{=}&  \nabla_i\nabla_j S[\varphi]  + \xi^p\nabla_p\nabla_i\nabla_j S[\varphi] + {\textstyle{\frac{1}{2!}}}\,\xi^p\xi^q\nabla_p\nabla_q\nabla_i\nabla_j S[\varphi] + \cdots \nn\\
&+& \mathrm{terms\;\,proportional\;\,to\;\,the\;\,Riemann\;\,tensor}.
\label{S2cov}
\eea}%
This expansion is valid only in the normal coordinate system, but since it appears as the argument of a logarithm in the one-loop effective action,
the tensor transformations required to take it to an arbitrary coordinate system will be cancelled by the Jacobian of the change of variables in the path integral measure. 
Then to prove the single-field dependence of the one loop effective action, or in other words,
to verify if promoting $\varphi\rightarrow\phi$ in \eqref{1loopeffactbg} will reproduce \eqref{1loopeffact}, one simply needs to see if the two expressions in Eqs.~(\ref{S2fluct}) and~(\ref{S2cov}) match.
This is clearly seen to be true if the connection is flat, so that the covariant derivatives commute.

The aim of the above discussion was to see explicitly the source of violation of single-field dependence due to an exponential splitting based on a non-flat connection.
Of course, taking advantage of the covariance of the effective action, there is a more general and even easier way to see its single-field dependence.
In fact one can move to a coordinate system where the flat connection vanishes. This leads to the linear splitting which gives rise to a single-field effective action.
Then, moving back to the original coordinates just changes the quantum-background dependence of the single total field and generates flat connections, i.e. changes ordinary derivatives to covariant derivatives. 

\subsection{Symmetries of the effective action} \label{sea}

The argument so far describes a quantization procedure which leads to a covariant and manifestly background-independent effective action.
It is now natural to ask whether a symmetry of the ultraviolet action is also preserved in the effective action.
To address this question let us take a look at the covariance relation \eqref{cov}, which tells us how the effective action changes under a general field transformation.
If we further assume that the transformation denoted by a prime is a symmetry of the ultraviolet action, i.e. $S_G[\phi']=S_G[\phi]$ or equivalently $G'=G$, then the covariance relation \eqref{cov} reduces to
\be \label{symm}
\Gamma_{G,f'}[\phi'] = \Gamma_{G,f}[\phi].
\ee
This is not exactly the symmetry property of the ultraviolet action because the flat connection present as an extra object in the effective action, has to be transformed as well.
The symmetry \eqref{symm} present in the infrared takes the same form as that in the ultraviolet action if and only if the connection coefficients are also invariant, i.e. $\Gamma'^k_{ij} = \Gamma^k_{ij}$, or more explicitly
\be 
\Gamma^k_{ij}(\phi') = \Gamma'^k_{ij}(\phi'), \qquad \Gamma^k_{ij}(\phi') = (\bar U^{-1})^l_i\, \bar U\left[\Gamma_l(\phi)+\partial_l\right]\bar U^{-1}, \qquad \bar U^i_j = \frac{\partial \phi'^i}{\partial \phi^j}.
\ee
Given a connection, the solution to this equation gives the set of transformations that, if present as symmetries in the ultraviolet, will also leave the effective action invariant. It is easy to solve this equation for a flat connection.
In such a case one can simply move to a coordinate system where the connection coefficients vanish $\Gamma^k_{ij}=0$, in which case the symmetry identity for the Christoffel symbols reduces to
\be 
(\bar U^{-1})^a_i\, \bar U\partial_a\bar U^{-1} = 0 \qquad \mathrm{or}\qquad \partial_a\bar U^{-1} = 0 \qquad \mathrm{or}\qquad \frac{\partial^2 \phi^i}{\partial \phi'^a\partial \phi'^b} =0.
\ee
This is simply saying that the symmetry transformations in the coordinates where the connection coefficients vanish must be linear. This gives us a criterion
for the preservation of the symmetries in the effective action: a symmetry is preserved in the effective action if and only if it is linearizable,
i.e. if there exists a choice of coordinates which transforms linearly under the symmetry, and if we choose our flat connection to vanish in this coordinate system. 
Of course the symmetry transformations and the connection coefficients transform accordingly when the fields are redefined. 

It is also possible to find a criterion for the cases where this linearization is possible. Let $H$ be a symmetry group acting on the manifold of fields $\mathcal{M}$.
A sufficient condition for the linearizability of the transformation follows from a lemma proved by Coleman, Wess and Zumino \cite{cwz}
which states that if at a point on $\mathcal{M}$ the group $H$ is preserved, i.e. if the group $H$ has a fixed point in $\mathcal{M}$,
then at least in a neighbourhood of this point there is a choice of coordinates on which the group $H$ acts linearly.
Now, to find the necessary condition, let us assume that there exists a set of coordinates which transforms linearly under the action of the group.
In this case, the point on the manifold corresponding to the zero values of the fields is a fixed point. But this point does not necessarily lie on the manifold.
For instance it might correspond to the infinite values of the original coordinates, as happens in a mapping of the cylinder to the plane.
So the necessary condition for the linearizability of the group action is that the field manifold have, or can be extended to have, a fixed point.
Interestingly, inspired by this, one can take a step further and extend the sufficient condition provided by the CWZ lemma to the case where the manifold
is obtained from the ones possessing a fixed point but with the fixed point removed from the target space\footnote{In fact one can even remove a region
including the fixed point, provided that the linearizable patch around the fixed point covers (at least part of) the manifold. With this assumption,
one can roughly state that the necessary and sufficient condition for the linearizability of the group transformation is that there exist a fixed point
of the symmetry group or that there is an extension of $\mathcal{M}$ on which $H$ smoothly extends to an action which possesses a fixed point.}.
In summary, a linearly transforming set of coordinates exists if there exists a fixed point of the symmetry group or if there is a one-point extension
of $\mathcal{M}$ on which $H$ smoothly extends to an action which leaves the included point invariant. 

The argument above, deals with the linearizability of the group action
and has nothing to do with the theory itself. In fact, it is important to distinguish between the linearizability of a theory and the linearizability of a group action.
The former requires, in addition to the existence of a linearly transforming coordinate system, that the Lagrangian have a good description in such coordinates.
It is also important to distinguish between a fixed point of the theory and a fixed point of the group action.
The former refers not only to an invariant point on the field space but also to the invariance of the Lagrangian at this fixed point.
With these definitions, a theory is linearizable if and only if it has a fixed point.

We just argued that whenever there is a coordinate system in which the symmetry group acts linearly on the fields, there is a flat connection which is invariant under the symmetry group.
This is the connection which vanishes in the linearly transforming coordinates. 

A theory defined on $\mathcal{M}\!-\!\{\mathrm{fixed\;\,point}\}$ does not necessarily have a good description in a coordinate system where the flat connection vanishes, but this does not invalidate our argument.
In fact we never need to refer to such a (Cartesian) coordinate system. This, for example, allows us to extend our argument to a general $O(N)$ invariant effective field theory of scalars,
whose field space is an $N$-dimensional manifold with arbitrary topology.
These include purely nonlinear sigma models such as $O(N)$ invariant theories defined on field spaces with the topology of a cylinder $\mathbb{R}\times S^{N-1}$.

An $O(N)$ invariant theory of $N$ scalars defined on a space with cylindrical topology provides an example of a nonlinearizable symmetry that is preserved in the effective action.
There are also examples of nonlinearizable symmetries such as an $O(N)$ invariant theory defined on the $(N-1)$--sphere, for which the symmetry is not preserved in the effective action.
However, generally, theories of the second type can be turned into the first type by adding extra (neutral) degrees of freedom, i.e. embedding them in higher dimensional spaces.
Choosing the extra degrees of freedom to be decoupled from the original ones, the physical content and universal critical properties of the original lower dimensional (target space) theory
can be extracted from the higher dimensional theory whose symmetry is preserved in the effective action. 

Finally we find it instructive to work out the infinitesimal form of the identity \eqref{symm}.
This consists of two pieces, one is the variation of the effective action $\delta_\phi\Gamma$ under an infinitesimal change $\bar\phi\rightarrow \bar\phi + \delta\bar\phi$ in the field,
and the other is $\delta_f\Gamma$ which is the result of the infinitesimal variation $f\rightarrow f - \delta\bar\phi \,\partial f$ induced by the corresponding change in the field.
The latter can be easily computed from the identity \eqref{alt_ea_equation}. The infinitesimal version of \eqref{symm} then takes the form
\be \label{symminfinitesimal}
\frac{\delta\Gamma}{\delta\bar\phi}\,(\partial f)^{\!-\!1}\!(f(\bar\phi))\,\langle (\partial f)(\phi)\,\delta\phi\rangle =0.
\ee
For the special case where $f$ is the identity function, that is where the splitting reduces to the standard linear one, the identity above takes the familiar form
\be \label{symminfinitesimalid}
\frac{\delta\Gamma}{\delta\bar\phi}\,\langle\delta\phi\rangle =0,
\ee
which is nothing but the usual Ward identity corresponding to the symmetry $\phi\rightarrow \phi + \delta\phi$.

\section{A couple of explicit examples}

We shall discuss here two simple examples.
The first is the case of a single scalar field theory for which a kinetic term can always be put in a canonical form by a suitable redefinition of the fields.
Then we briefly discuss some aspects of the linear $O(2)$ model with a flat target space wherein some extra features of the background-fluctuation splitting can be shown in this covariant formalism. 

\subsection{The case of one field} \label{ssec.1f}
In order to illustrate the approach in the simplest terms
let us take as an example the following sigma model with a single scalar field
\be \label{stof} 
S[\phi] = \int_x \left[ {\textstyle{\frac{1}{2}}}\,J(\phi)\,\partial_\mu\phi\partial^\mu\phi - V(\phi) \right].
\ee
where we consider here, as well as in what follows, a space-time with a Lorentzian (mostly minus) signature. The action \eqref{stof} is the most general scalar theory with at most two derivatives. 

Following our prescription, we use the background-field method with the exponential splitting based on a flat connection, denoted by $\Gamma$.
This should not be confused with the effective action. A connection in a one dimensional space is always flat, so $\Gamma$ can be written as
\be 
\Gamma = - (f^{-1})''(f')^2
\ee
for some function $f$. The expansion for flat splitting therefore takes the form
\be \label{glspexp} 
\phi(\varphi,\xi) = (f^{-1})\left(f(\varphi)+ \partial f\xi\right) = \varphi + \xi - {\textstyle{\frac{1}{2}}}\Gamma\,\xi^2 + \cdots
\ee
To find the one-loop effective action \eqref{1loopea}, we need to compute $S^{(2)}[\varphi]$, which is the second $\xi$-derivative of $S[\phi]$ at $\xi=0$,
or $\nabla^2S[\varphi]=S''[\varphi]-\Gamma S'[\varphi]$, where $\nabla$ is the covariant derivative with respect to the connection $\Gamma$. For example we have 
\be 
\nabla_\mu = \partial_\mu + \partial_\mu\varphi\, \Gamma, \quad \nabla J = J' - 2\Gamma J, \quad \nabla^2 J = (\nabla J)' - 3\Gamma \nabla J, \quad \nabla V = V', 
\quad \nabla^2 V = V'' - \Gamma V'.
\ee
The two quantities are of course equal. Without referring to the flatness of the connection, these are found to be 
\be 
S^{(2)}[\varphi] = -J\,\nabla\!_\mu\!\nabla^\mu - \nabla J \partial_\mu\varphi\nabla^\mu
- \nabla J \nabla_\mu\partial_\mu\varphi -{\textstyle{\frac{1}{2}}}\nabla^2 J\, \partial_\mu\varphi\partial^\mu\varphi - \nabla^2 V(\varphi)
\ee
\be
S''[\varphi] - \Gamma(\varphi)S'[\varphi] = -J\partial^2 - J'\partial_\mu\varphi\partial^\mu - {\textstyle{\frac{1}{2}}}J''\partial_\mu\varphi\partial^\mu\varphi + {\textstyle{\frac{1}{2}}}\Gamma J' \partial_\mu\varphi\partial^\mu\varphi -J' \partial^2 \varphi+ \Gamma J \partial^2\varphi - V'' + \Gamma V',
\ee
where $S^{(2)}[\varphi]$ is the second $\xi$-derivative of $S[\phi]$ at $\xi=0$, and. 
One can easily check that the above two expressions are indeed the same.
To be concrete let us stick to four space-time dimensions. For simplicity we restrict to the case of a constant background, which only gives the one-loop effective potential
{\setlength\arraycolsep{2pt}
\bea 
\Gamma^{1-\mathrm{loop}}_{\!f}[\phi] &=& S[\phi] + {\textstyle{\frac{i}{2}}}\mathrm{Tr}\log\left[S,_{ij}-\Gamma^k_{ij}\,S,_k\right] = S[\phi] + {\textstyle{\frac{i}{2}}}\mathrm{Tr}\log\left[-J(\phi)\,\partial^2 - \nabla^2_{\!f} V(\phi)\right] \nn\\
&=& S[\phi] + {\textstyle{\frac{i}{2}}}\mathrm{Tr}\log\left[-\partial^2 - \nabla^2_{\!f} V(\phi)/J(\phi)\right] + {\textstyle{\frac{1}{2}}}\mathrm{Tr}\log\left[J(\phi)\right].
\eea}%
To proceed with the computation, we adopt dimensional regularization $d= 4-2\epsilon$. The last term therefore vanishes. Renaming the quantity $\nabla^2_{\!f} V(\phi)/J(\phi) = F(\phi)$ the second term becomes
\be 
{\textstyle{\frac{i}{2}}}\int_p\log\left[p^2 + F(\phi)\right].
\ee
This can be computed, restricting to a constant field and for example using equation (11.72) of \cite{peskin}:
{\setlength\arraycolsep{2pt}
\bea \label{intdiv}
{\textstyle{\frac{i}{2}}}\,\mu^{2\epsilon}\!\int_p\log\left[p^2 + F(\phi)\right] &=& {\textstyle{\frac{1}{2}}}\mu^{2\epsilon} \frac{\Gamma(-d/2)}{(4\pi)^{d/2}} F^{d/2} \nn\\
&=& \frac{F^2}{2(4\pi)^2}\left(\frac{1}{2\epsilon} - \gamma +\frac{3}{2} + \mathcal{O}(\epsilon)\right) \left(1+ \epsilon \log(4\pi\mu^2) + \mathcal{O}(\epsilon)\right) \left(1-\epsilon\log F + \mathcal{O}(\epsilon)\right) \nn\\
&=& \frac{F^2}{2(4\pi)^2}\left(\frac{1}{2\epsilon} - \gamma +\frac{3}{2}- \frac{1}{2} \log\frac{F}{4\pi\mu^2} + \mathcal{O}(\epsilon)\right) .
\eea}%
For the more general case of a space-time dependent field the calculation of the divergent term in dimensional regularization will be discussed in Section \eqref{sec.nlsm}. In the $\overline{\mathrm{MS}}$ scheme, the renormalized finite part becomes
\be 
{\textstyle{\frac{i}{2}}}\int_p\log\left[p^2 + F(\phi)\right] \stackrel{\mathrm{ren}}{=} -\frac{F^2}{4(4\pi)^2}\log\frac{F}{4\pi\mu^2} ,
\ee
so finally the one-loop effective potential reads
\be \label{1lep}
V^{1-\mathrm{loop}}_{\!J,V,f}(\phi) = V(\phi) + \frac{(\nabla^2_{\!f} V(\phi))^2}{4(4\pi)^2J^2(\phi)}\log\frac{\nabla^2_{\!f} V(\phi)}{4\pi\mu^2 J(\phi)} .
\ee
Notice that the quantity $F$ is a scalar, so the above equation is covariant as expected. This is in the sense that
\be 
V^{1-\mathrm{loop}}_{\!J',V',f'}(\phi') = V^{1-\mathrm{loop}}_{\!J,V,f}(\phi), 
\label{cov_one_field}
\ee
where 
\be 
\phi' = h(\phi), \; J'(\phi') = (\partial h^{-1})^2J(\phi), \;V'(\phi') = V(\phi), \; f'(\phi') = f(\phi).
\ee
Moreover the one-loop effective action is manifestly single field dependent and therefore one can compute it just for the
background field and then trivially reconstruct the full dependence to all orders in the fluctuations. 
This means that with the above computation one can obtain the effective potential contribution to all the 1PI off-shell vertices at one loop.

In the above analysis we have adopted dimensional regularization and $\overline{\mathrm{MS}}$ which we have found to be the most convenient scheme to study perturbative renormalization in the functional form. However, one can regularize the theory in other ways, such as cutting off the Euclidean momenta explicitly at $p=\Lambda$. In such a case, for instance, the integral in \eqref{intdiv} evaluates to 
\be 
{\textstyle{\frac{1}{2}}}\int_{|p|\leq\Lambda}\hspace{-5pt}\log\left[p^2 + F(\phi)\right] = \frac{1}{2(4\pi)^2}\left[F\Lambda^2+\frac{1}{2}F^2\left(\log\frac{F}{\Lambda^2}-\frac{1}{2}\right)\right]+\mathcal{O}(1/\Lambda^2).
\ee 
where we have dropped the $\mathcal{O}(\Lambda^4)$ field-independent terms. In order to remove the divergences one needs to impose some RG conditions which, to guarantee the covariance of the resulting effective action, must be covariant themselves. For simplicity of discussion let us restrict to renormalizable theories, i.e. those which involve operators of dimension no more than four in some parametrization, and further require $\mathbb{Z}_2$ symmetry in such parametrization. In this case one needs only two RG conditions which can naturally be taken to be $\nabla^2_{\!f} V_{eff}(0)=0$, which fixes the mass to zero, and $(\nabla^2_{\!f})^2 V_{eff}(M)=(\nabla^2_{\!f})^2 V(0)$, where $M$ is some renormalization scale. These are simply covariant extensions of the RG conditions used in \cite{Coleman:1973jx}. With these conditions the one-loop effective action is found to take the following simple form which is both covariant and single-field dependent
\be 
V^{1-\mathrm{loop}}(\phi) = V(\phi) + \frac{F^2(\phi)}{4(4\pi)^2}\left(\log\frac{F(\phi)}{F(M)}-\frac{25}{6}\right).
\ee
Let us summarize the main advantages of this procedure. 
The one-loop effective potential can be computed with the background-field method using the nonlinear quantum-background splitting introduced in Section $3$. 
The result \eqref{1lep} is manifestly background independent, in the sense that it depends on the total field only. 
Notice that in Eq.~\eqref{1lep} the total field is given by \eqref{fs} with the quantum field replaced by its average.
Moreover Eq.~\eqref{1lep} transforms covariantly under field redefinitions of the total field. 
One might wonder if, through the covariant derivative $\nabla_f$ there is an extra dependence on the function $f$ in Eq.~\eqref{1lep}. 
In fact, this function must be fixed and in particular one can relate it to the function $J$. 
In the model considered here there is a natural choice for $f$. We know that a redefinition $\phi\rightarrow h(\phi)$, 
where $h'(\phi) = J^{1/2}(\phi)$, brings the field into canonical form. 
It is natural to assign the vanishing connection to the canonical coordinates $h$. 
With this choice, the connection in the original $\phi$ coordinates satisfies $f'(\phi)=J^{1/2}(\phi)$. 

\subsection{Flat linear $O(2)$ model, in Cartesian and polar coordinates} 
As a second example we consider here the two scalar-field linear $O(2)$ model in flat field space, 
first discussed by Kunstatter~\cite{kunstatter}, whose Lagrangian in Cartesian coordinates is given by
\be \label{o2c} 
\mathcal{L} = {\textstyle{\frac{1}{2}}}\,\partial_\mu\phi^i\partial^\mu\phi^i - V(\phi^2), \qquad \phi^2 = \phi^i\phi^i.
\ee
We start with performing a standard one-loop computation of the effective action in this frame. One needs the second derivative of the action 
{\setlength\arraycolsep{2pt}
\bea
S,_{ij} 
&=& -\square\,\delta_{ij} - 2\delta_{ij} V'(\phi^2) - 4\phi^i\phi^j V''(\phi^2) \nn\\
&=& \left(-\square - 2V'\right)(P_\perp)_{ij} + \left(-\square - 2V'-4\phi^2 V''\right)(P_L)_{ij} \,,
\eea}%
where in the last equation we have introduced the projectors
\be
(P_\perp)_{ij} \equiv \delta_{ij} -\phi^i\phi^j/\phi^2, \; (P_L)_{ij} \equiv \phi^i\phi^j/\phi^2 \,.
\ee
Then one finds trivially
\be 
\mathrm{Tr}\log S,_{ij} = \mathrm{Tr}\log\left(-\square - 2V'\right) + \mathrm{Tr}\log\left(-\square - 2V'- 4\phi^2 V''\right).
\ee
In four space-time dimensions, and restricting to a renormalizable theory, the divergent part of the one-loop effective action in dimensional regularization with $d=4-2\epsilon$ is
\be 
\mathrm{div}\, {\textstyle{\frac{i}{2}}}\mathrm{Tr}\log S,_{ij}  = \frac{V'^2+(V'+2\phi^2 V'')^2}{2(4\pi)^2\epsilon}
\ee 
and in the $\overline{\mathrm{MS}}$ scheme the one loop effective potential becomes
{\setlength\arraycolsep{2pt}
\bea  \label{1lepc}
{\textstyle V_{eff} }&=& V+\frac{V'^2}{(4\pi)^2}\log\frac{V'}{2\pi\mu^2}+ \frac{(V'+2\phi^2 V'')^2}{(4\pi)^2}\log\frac{V'+2\phi^2 V''}{2\pi\mu^2}, \qquad V(x) = \frac{\lambda}{4!} x^2, \nn\\
&=&  \frac{\lambda}{4!} \phi^4+\frac{\lambda^2\phi^4}{144(4\pi)^2}\log\frac{\lambda\phi^2}{24\pi\mu^2} +\frac{9\lambda^2\phi^4}{144(4\pi)^2}\log\frac{3\lambda^2\phi^2}{24\pi\mu^2} \nn\\
&=&  \frac{\lambda}{4!} \phi^4+\frac{\lambda^2\phi^4}{4(4\pi)^2}\,\frac{5}{18}\,\log\frac{\lambda\phi^2}{24\pi\mu^2} +\frac{9\lambda^2\phi^4}{144(4\pi)^2}\log 3.
\eea}%
We note that, since the defining theory is linear, in such a frame one can consider a simple linear background-fluctuation splitting which is respected by the quantization procedure so that the result can be considered single field dependent.

We now repeat the computation in polar coordinates. Let us first define the mapping among the two charts of the target manifold: 
\be \label{cp} 
\phi^1 =\rho \sin\theta =  f^1(\rho,\theta), \qquad \phi^2 = \rho \cos\theta  = f^2(\rho,\theta).
\ee
Rewriting \eqref{o2c} in terms of these new fields give the Lagrangian
\be 
\mathcal{L} = {\textstyle{\frac{1}{2}}}\,\partial_\mu\rho\partial^\mu\rho + {\textstyle{\frac{1}{2}}}\rho^2\,\partial_\mu\theta\partial^\mu\theta - V(\rho).
\ee
To find the one-loop effective potential, according to the general formula, we need to know the connection. This is given by 
\be 
\Gamma^k_{ij} = [(\partial f)^{-1}]^k_a\, \partial_i\partial_j f^a
\ee
where the functions $f^i$ are defined in \eqref{cp}. The matrix in the brackets is the inverse of 
\be 
\ba{r} 
\ba{cc}  \rho & \hspace{8mm} \theta\phantom{\cos\theta} \ea \\
\left[\partial f\right]_{ij} = \partial_j f^i, \qquad \partial f = \ba{c} 1 \\ 2  \ea \!\! \left(\ba{cc} \sin\theta & \rho\cos\theta \\ \cos\theta & -\rho\sin\theta \ea\right) .
\ea
\ee
Also, regarding the factor on the right as two matrices $[\partial^2f^a]_{ij} = \partial_i\partial_jf^a$, $a=1,2$
\be 
\ba{c}
\ba{r} 
\ba{cc}  \rho & \hspace{8mm} \theta \phantom{\cos\theta} \ea \\
\partial^2 f^1 = \ba{c} \rho \\ \theta  \ea \!\! \left(\ba{cc} 0 & \cos\theta \\ \cos\theta & -\rho\sin\theta \ea\right) 
\ea \qquad
\ba{r} 
\ba{cc}  \rho & \hspace{1cm} \theta \phantom{\cos\theta} \ea \\
\partial^2 f^2 = \ba{c} \rho \\ \theta  \ea \!\! \left(\ba{cc} 0 & -\sin\theta \\ -\sin\theta & -\rho\cos\theta \ea\right) .
\ea
\ea
\ee
The connection, also considered as a matrix $[\Gamma^a]_{ij} = \Gamma^a_{ij}$, can be expressed as
\be 
\left(\ba{c} \Gamma^\rho \\ \Gamma^\theta \ea\right) = \left(\ba{cc} \sin\theta & \cos\theta \\ \cos\theta /\rho & -\sin\theta/\rho \ea\right) \left(\ba{c} \partial^2\! f^1 \\ \partial^2\! f^2 \ea\right),
\ee
explicitly, the components of the connection are found to be
\be \label{connpolar}
\Gamma^\rho  =  \left(\ba{cc} 0 & 0 \\ 0 & -\rho \ea\right) \,,
\quad
\Gamma^\theta =  \left(\ba{cc} 0 & 1/\rho \\ 1/\rho & 0 \ea\right).
\ee
So the only non zero components of the Christoffel symbols are $\Gamma^\rho_{\theta\theta} = -\rho$ and $\Gamma^\theta_{\rho\theta} = 1/\rho$.
We stress that in the polar coordinates, in order to get an effective potential which is simply a result of expressing \eqref{1lepc} in terms of $\rho,\theta$, a linear splitting
cannot be used. Instead, one needs a non-linear splitting according to $\phi^i(\varphi,\xi) = (f^{-1})^i\left(f(\varphi)+ \partial f\xi\right)$, with the function $f^i$ defined in \eqref{cp}.
Denoting with $\rho_0$ and $\theta_0$ the background fields and with $\xi_\rho$ and $\xi_\theta$ the fluctuations, the non linear splitting is given by
\bea
\rho&=&\sqrt{(\rho_0+\xi_\rho)^2+\rho_0^2\,\xi_\theta^2}\,,\nn\\
\theta&=&\arctan{ \frac{\sin \theta_0 \left(\rho_0+\xi_\rho\right)+\rho_0 \cos \theta_0\, \xi_\theta}{\cos \theta_0 \left(\rho_0+\xi_\rho\right)-\rho_0 \sin\theta_0 \,\xi_\theta}}.
\label{polar_split}
\eea
One can check that expanding in powers of the fluctuations the expressions for the exponential splitting in terms of the connection are recovered
\bea
\rho&=&\rho_0+\xi_\rho+\frac{1}{2}\rho_0\, \xi_\theta^2-\frac{1}{2}\xi_\rho \xi_\theta^2+\cdots\nn\\
\theta&=&\theta_0+\xi_\theta-\frac{1}{\rho_0} \xi_\rho \xi_\theta+\frac{1}{\rho_0^2}\xi_\rho^2 \xi_\theta-\frac{1}{3}\xi_\theta^3+\cdots
\eea
In fact, one could have read off the components of the connection this way, by looking at the coefficients of the quadratic terms.
We can now compute the second variation of the action in polar coordinates
\be 
S^{(2)} = -\left(\ba{cc} 1 & 0 \\ 0 & \rho^2 \ea\right)\square - \left(\ba{cc} V''(\rho) & 0 \\ 0 & 0 \ea\right), \qquad S,_\rho = - V'(\rho), \quad S,_\theta = 0,
\ee
Using the connection \eqref{connpolar}, the second covariant derivative (at the background level) is 
\be 
S^{(2)} - \Gamma^\rho S,_\rho = -\left(\ba{cc} 1 & 0 \\ 0 & \rho^2 \ea\right)\square - \left(\ba{cc} V''(\rho) & 0 \\ 0 & \rho V'(\rho) \ea\right) = \left(\ba{cc} 1 & 0 \\ 0 & \rho^2 \ea\right)\left[-\square - \left(\ba{cc} V''(\rho) & 0 \\ 0 &  V'(\rho)/\rho \ea\right)\right],
\ee
and finally the expression for the one-loop effective potential is found by computing the trace 
\be 
{\textstyle{\frac{i}{2}}}\mathrm{Tr}\log\left(S^{(2)} - \Gamma^\rho S,_\rho\right) = {\textstyle{\frac{i}{2}}}\mathrm{Tr}\log\left(-\square - V''(\rho)\right) + {\textstyle{\frac{i}{2}}}\mathrm{Tr}\log\left(-\square - V'(\rho)/\rho\right) + \mathrm{Tr}\log\rho.
\label{SeffO2}
\ee
Therefore, if we rename the potential in the Cartesian theory as $\tilde V(\phi^i\phi^i)$, it is readily seen that \eqref{SeffO2} is nothing but the one-loop effective potential in Cartesian coordinates rewritten in terms of the polar fields ($\phi^i\phi^i=\rho^2$)
\be 
V(\rho) = \tilde V(\rho^2), \quad\Rightarrow\quad V'(\rho) = 2\rho\tilde V'(\rho^2), \quad V''(\rho) = 2\tilde V'(\rho^2) + 4\rho^2 \tilde V''(\rho^2).
\ee
We see that employing the nonlinear flat splitting the off-shell effective potential is indeed fully covariant, as expected from the discussions in Section 2.3.
The results presented here are also valid not only at the background level but at all orders of the fluctuation field.
For this linear model which has a flat target space the covariant method of Vilkovisky matches our prescription and enjoys the single-field property.
We stress here that in our approach the single-field property continues to hold for models with non-flat field spaces. In the next section we generalize the discussion to such models.

\section{The nonlinear sigma model} \label{sec.nlsm}

So far we have been dealing with examples that are essentially in flat field space. In these examples the flat connection used to construct
the exponential map coincides with that compatible with the metric of the theory. For a theory on curved space this is no longer true.
Let us now consider the most general effective field theory of scalars at second order in the derivative expansion. 
\be \label{nlsm}
S[\phi] =\int_x  \left[{\textstyle{\frac{1}{2}}}\,g_{ij}(\phi)\,\partial_\mu\phi^i\partial^\mu\phi^j - V(\phi) \right].
\ee
This is characterized by a metric $g_{ij}(\phi)$ and a potential $V(\phi)$. We would like to compute, in dimensional regularization,
the one-loop counter-terms required to cancel the divergences in the effective action. Following our prescription to construct a covariant single-field effective action, we use the background-field method with the exponential splitting with a flat connection 
\be \label{glspexp} 
\phi^i(\varphi,\xi) = (f^{-1})^i\left(f(\varphi)+ \partial f\xi\right) = \varphi^i + \xi^i - {\textstyle{\frac{1}{2}}}\Gamma^i_{pq}\,\xi^p\xi^q + \cdots, \qquad 
\Gamma^i_{pq} = - (f^{-1})^i\!\!,_{ab}f^a\!\!\!,_p f^b\!\!\!,_q
\ee
To find the one-loop effective action \eqref{1loopea}, we need to compute the second variation of the action with respect to the fluctuation fields.
For notational convenience let us stick to the background computation but keep in mind that one can promote the background field to the total field.
At the background level, this second variation is nothing but the second covariant derivative $\nabla$ of the action, with the covariant derivative being compatible with our flat connection $\Gamma^k_{ij}$.
However, it is useful to write this in terms of the covariant derivative $\tilde\nabla$ compatible with the metric $g_{ij}$ of the theory
\be
\tilde \Gamma^k_{ij}={\textstyle{\frac{1}{2}}} g^{kl}\left(g_{lj,i}+g_{li,j} -g_{ji,l}\right).
\ee
These are simply related as follows
\be 
S_{;ij} = 
\nabla_i\nabla_j S = S,_{ij} - \Gamma^k_{ij} S,_k = S,_{ij} - \tilde{\Gamma}^k_{ij} S,_k + \delta\Gamma^k_{ij} S,_k = \tilde\nabla_i\tilde\nabla_j S + \delta\Gamma^k_{ij} S,_k
\ee
where 
\be
\delta \Gamma^i_{kj} =\tilde \Gamma^i_{kj}-\Gamma^i_{kj}={\textstyle{\frac{1}{2}}} g^{il}\left(\nabla_k g_{lj}+ \nabla_j g_{lk} - \nabla_l g_{jk}\right).
\ee
It is convenient to raise the indices with the metric $g_{ij}$
so that in the expression for the one-loop effective action the quantity
\be \label{gScompact}
g^{il} S_{;lj} = g^{il}\tilde\nabla_l\tilde\nabla_j S + g^{il}\delta\Gamma^k_{lj} S,_k
\ee
appears in the argument of the logarithm. The first term in \eqref{gScompact} is the usual quantity used in the covariant background-field method.
The second term is proportional to the equations of motion. It does not affect physical quantities, but it makes the effective action single-field dependent. This dependence is in fact the same as the one in Eq.~\eqref{fs}.
Of course one could have used $\nabla_i\nabla_j S = S,_{ij} - \Gamma^k_{ij}S,_k$. A direct computation gives
\be 
S,_i[\varphi] = \frac{\delta S}{\delta \varphi^i(x_i)}={\textstyle{\frac{1}{2}}}\,\partial_i g_{pq}\,\partial_\mu\varphi^p\partial^\mu\varphi^q - \partial_k g_{iq}\,\partial_\mu\varphi^q\partial^\mu\varphi^k - g_{iq}\partial^2\varphi^q - V,_i(\varphi)
\ee
{\setlength\arraycolsep{2pt}
\bea 
S,_{ij}[\varphi] &=&\frac{\delta^{2} S}{\delta \varphi^j(x_j)\delta \varphi^i(x_i)} = {\textstyle{\frac{1}{2}}}\,\partial_i\partial_j g_{pq}\,\partial_\mu\varphi^p\partial^\mu\varphi^q - \partial_j\partial_q g_{ip}\,\partial_\mu\varphi^p\partial^\mu\varphi^q -  g_{ij}\partial^2 \nn\\
&-&  \partial_k g_{ij}\,\partial_\mu\varphi^k\partial^\mu +\partial_i g_{jq}\,\partial_\mu\varphi^q\partial^\mu  - \partial_j g_{iq}\,\partial_\mu\varphi^q\partial^\mu - \partial_j g_{iq}\,\partial^2\varphi^q - V,_{ij}(\varphi)
\label{S2}
\eea}%
where the latter is symmetric under the exchange $(i,x_i) \leftrightarrow (j,x_j)$.
To see explicitly this fact it helps to rewrite the second, third and fourth terms in the second line of Eq.~(\ref{S2}) reintroducing the space time integral and the delta distributions in order to perform the integrations by parts.
Let us notice that in the operator $S,_{ij}[\varphi]$ there are single derivative terms
which are eliminated passing to a description with the covariant derivative $\tilde \nabla$:
this is the main reason to adopt such a choice. Indeed the first term in \eqref{gScompact} has a simple expression in terms of $g_{ij}$ and $V$
\be \label{gtS}
g^{il}\tilde\nabla_l\tilde\nabla_j S = - \delta^i_j\tilde\nabla_\mu\tilde\nabla^\mu - \tilde R^i_{pjq}\partial_\mu\varphi^p\partial^\mu\varphi^q- g^{il}\tilde\nabla_l\tilde\nabla_j V
\ee
where indices are raised and lowered by $g_{ij}$. In our convention the Riemann tensor is defined as
\be
\tilde R_{j q\phantom{i}p}^{\phantom{jq}i} =  \partial_j \tilde{\Gamma}^i_{qp} - \partial_q\tilde\Gamma^i_{jp} + \tilde{\Gamma}^i_{jl}\tilde{\Gamma}^l_{qp}-\tilde{\Gamma}^i_{ql}\tilde{\Gamma}^l_{jp} \nn.
\ee
Using \eqref{gtS} one can write \eqref{gScompact} as 
\be \label{gs}
g^{i l} S_{; l j}= - \delta^i_j\tilde\nabla_\mu\tilde\nabla^\mu-Q^i_j
\ee
where we have defined
\be  \label{Qgeneral}
Q^i_j = \tilde R^i_{pjq}\partial_\mu\varphi^p\partial^\mu\varphi^q + g^{il}\tilde\nabla_l\tilde\nabla_j V + g^{ip}g_{kq}\delta\Gamma^k_{pj}\tilde\nabla_\mu \partial^\mu\varphi^q + g^{ip}\delta\Gamma^k_{pj}V,_k .
\ee
Recall also that
\be 
g^{il}\tilde\nabla_l\tilde\nabla_j V + g^{ip}\delta\Gamma^k_{pj}V,_k = g^{il}\nabla_l\nabla_j V .
\ee
We now compute the divergent part of the one-loop effective action in dimensional regularization.
We are especially interested in two and four space-time dimensions. Denoting the space-time dimension by $n$, in dimensional regularization we choose $d=n-2\epsilon$.
The general formula giving the (logarithmic) divergence in Minkowski space can be obtained with Heat-Kernel techniques~\cite{dewitt1964}
\be \label{div}
\mathrm{div} \,\frac{i}{2}\mathrm{Tr}\log[\tilde\nabla_\mu\tilde\nabla^\mu + Q] = \frac{1}{2\epsilon(4\pi)^{n/2}}\int_x \mathrm{tr}\,a_{n/2}.
\ee
In $n=2$ we just need $a_1=-Q$ so that
\be \label{oneloopdiv2}
\mathrm{div} \,\frac{i}{2}\mathrm{Tr}\log[\tilde\nabla_\mu\tilde\nabla^\mu + Q] = - \frac{1}{8\pi \epsilon}\int_x \mathrm{tr}\left[Q\right].
\ee
In $n=4$ we need to know $a_2$, which is 
{\setlength\arraycolsep{2pt}
\bea 
\label{a2}
a_2 &=& \frac{1}{2}Q^2 + \frac{1}{6}[\tilde\nabla^\mu,[\tilde\nabla_\mu, Q]] + \frac{1}{12}[\tilde\nabla^\mu,\tilde\nabla^\nu][\tilde\nabla_\mu,\tilde\nabla_\nu], \qquad [\tilde\nabla_\mu,\tilde\nabla_\nu]^a_b v^b= \partial_\mu\varphi^p\partial_\nu\varphi^q \tilde R_{pq\phantom{a}b}^{\phantom{pq}a} v^b  \nn\\
&=& \frac{1}{2}Q^2 + \frac{1}{6}\tilde\nabla^\mu\tilde\nabla_\mu Q + \frac{1}{12}\,\partial^\mu\varphi^p\partial^\nu\varphi^q \partial_\mu\varphi^k\partial_\nu\varphi^l\, \tilde R_{pq\phantom{a}c}^{\phantom{pq}a} \tilde R_{kl\phantom{c}b}^{\phantom{kl}c}.
\eea}%
The divergent part of the one-loop effective action in four dimensions is therefore given by 
\be \label{oneloopdiv4}
\mathrm{div} \,\frac{i}{2}\mathrm{Tr}\log[\tilde\nabla_\mu\tilde\nabla^\mu + Q] = \frac{1}{4\epsilon(4\pi)^{2}}\int_x \mathrm{tr}\left[Q^2 + \frac{1}{6}\,\tilde R_{\mu\nu\phantom{a}k}^{\phantom{\mu\nu}i} \tilde R^{\mu\nu\, k}_{\phantom{\mu\nu i}j}\right], 
\ee
where 
\be 
\tilde R_{\mu\nu\phantom{a}j}^{\phantom{\mu\nu}i} = \partial_\mu\varphi^p\partial_\nu\varphi^q \tilde R_{pq\phantom{i}j}^{\phantom{pq}i}.
\ee
Compared to the $n=2$ case, the expression for one loop divergences in dimension $n=4$ is more involved. In particular one needs to compute the square of the matrix $Q^i_j$.
For later use we report it here
{\setlength\arraycolsep{2pt}
\bea \label{Q2general}
Q^i_j Q^j_i &=& \tilde R^i_{pjq}\tilde R^j_{kil}\partial_\mu\varphi^p\partial^\mu\varphi^q \partial_\nu\varphi^k\partial^\nu\varphi^l + 2\tilde\nabla^i\tilde\nabla_j V \tilde R^j_{kil}\partial_\mu\varphi^k\partial^\mu\varphi^l + \tilde\nabla^i\tilde\nabla_j V \tilde\nabla^j\tilde\nabla_i V \nn\\
&+& g^{ip}g_{kq}\delta\Gamma^k_{pj}\tilde\nabla_\mu \partial^\mu\varphi^q  
\tilde R^j_{ris}\partial_\nu\varphi^r\partial^\nu\varphi^s + g^{ip}\delta\Gamma^k_{pj}V,_k \tilde R^j_{ris}\partial_\nu\varphi^r\partial^\nu\varphi^s \nn\\
&+& g^{ip}g_{kq}\delta\Gamma^k_{pj}\tilde\nabla_\mu \partial^\mu\varphi^q 
\,\tilde\nabla^j\tilde\nabla_i V + g^{ip}\delta\Gamma^k_{pj}V,_k \tilde\nabla^j\tilde\nabla_i V + g^{ip}g^{jr}\delta\Gamma^k_{pj}\delta\Gamma^l_{ri}V,_k V,_l\nn\\
&+& 
g^{ip}g^{jr}g_{kq}\,g_{ls}\,\delta\Gamma^k_{pj}\delta\Gamma^l_{ri}\tilde\nabla_\mu \partial^\mu\varphi^q\tilde\nabla_\nu \partial^\nu\varphi^s
+2 g^{ip}g^{jr}g_{kq}\delta\Gamma^k_{pj}\delta\Gamma^l_{ri}\tilde\nabla_\mu \partial^\mu\varphi^q V,_l 
\eea}%
The first two terms in \eqref{Qgeneral} and the first line in \eqref{Q2general} give the usual terms found by expanding with the connection compatible with $g_{ij}$.
The rest are deviations from that due to our different choice of connection $\Gamma^k_{ij}$ for the exponential expansion.
They vanish when $\delta\Gamma^k_{ij} =\tilde\Gamma^k_{ij}-\Gamma^k_{ij}=0 $. Since $\Gamma^k_{ij}$ is chosen to be flat,
these extra contributions will make the expression for the one-loop divergence of the effective action valid to all orders of the fluctuation field,
while without them equations \eqref{oneloopdiv2} and \eqref{oneloopdiv4} are valid only at the background level, i.e. when the fluctuations are set to zero $\xi^i=0$. 

Despite the general discussions of Section~2, it might still be useful to perform an explicit check of the single-field dependence.
In appendix B we discuss a more explicit version of the general analysis given in Section~\ref{sfd} with a particular focus on the effective potential.

\section{$O(N)$ Effective Field Theory}

In this section we apply the ideas developed in previous sections to an effective field theory of $N$ scalars, with $O(N)$ symmetry. 
Instead of working in a completely general coordinate system ,we choose to do the analysis in the ``polar'' coordinates, which makes the $O(N)$ symmetry manifest.
In the polar coordinate system, $N-1$ fields $\chi^\alpha$, interpreted as ``angular'' fields, are used to parametrize the orbits of $O(N)$, and the extra ``radial'' field $h$, which is invariant under $O(N)$,
parametrizes different orbits. The angular fields transform nonlinearly under $O(N)$, and are left unspecified throughout the section. 

We emphasise that the topology of field space is also left arbitrary. The particular case of $N=4$ leads to an effective field theory of the Higgs scalar with custodial symmetry called Higgs effective field theory (HEFT) in \cite{manohar_1511}.

Our goal will be to compute the one loop divergences of the effective action in dimensional regularization
and discuss its different properties including covariance, single-field dependence, and symmetry properties, expected from the discussions in Section 2.
 
To be specific, let us consider the model defined by the following action
which contains derivatives up to second order
\be \label{oeft}
\mathcal{L} = \frac{1}{2}\partial_\mu h\partial^\mu h + \frac{1}{2} F^2(h)\, g_{\alpha\beta}(\chi)\partial_\mu\chi^\alpha\partial^\mu\chi^\beta - V(h),
\ee
where $g_{\alpha\beta}$ is the $O(N)$ invariant metric on the unit $(N-1)$--sphere. It is convenient to
introduce in the $(h,\chi^\alpha)$ space,  the $O(N)$ invariant metric
\be \label{metric}
G_{ij} = \left(\ba{cc} 1 & 0 \\ 0 & F^2(h)g_{\alpha\beta}(\chi) \ea\right),
\ee
which characterizes the kinetic term in \eqref{oeft}. Clearly the special choice $F(h)=h$ leads to a flat metric which is simply a polar reparametrization of the identity metric in the linearly transforming coordinate system. According to our prescription, in order to have a covariant single-field effective action which inherits the $O(N)$ symmetry of \eqref{oeft}, this is the flat metric that has to be used to construct the geodesics defining the exponential splitting.

The nonzero components of the connection $\tilde\Gamma^k_{ij}$ compatible with \eqref{metric} are given by
\be 
\tilde\Gamma^0_{\alpha\beta} = -FF' g_{\alpha\beta}, \qquad \tilde\Gamma^\alpha_{0\beta} = (F'\!/F)\,\delta^\alpha_\beta, \qquad \tilde\Gamma^\delta_{\alpha\beta} = (\Gamma_g)^\delta_{\alpha\beta}
\ee
where $(\Gamma_g)^\delta_{\alpha\beta}$ is the connection compatible with the metric $g_{\alpha\beta}$. The difference between this connection and the flat one $\Gamma^k_{ij}$ given by $F(h)=h$ appears in the expression \eqref{gScompact} for the second derivative of the action. The nonzero components of $\delta\tilde\Gamma^k_{ij} = \tilde\Gamma^k_{ij} - \Gamma^k_{ij}$ are given by
\be  \label{DelGam}
\delta\tilde\Gamma^0_{\alpha\beta} = -(FF'-h) g_{\alpha\beta}, \qquad \delta\tilde\Gamma^\alpha_{0\beta} = (F'\!/F- 1/h)\,\delta^\alpha_\beta = ((F'h-F)/Fh)\,\delta^\alpha_\beta.
\ee
Also, the Riemann and Ricci tensors associated with \eqref{metric} appear in the expression for the one-loop divergences of the effective action in two and four dimensions. For later use, we collect the non zero components here~\cite{PS}
{\setlength\arraycolsep{2pt}
\be \label{Riemann}
\ba{rcl} 
\tilde R_{\alpha\gamma\beta\delta} &=& (1-F'^2)F^2 (g_{\alpha\beta}g_{\gamma\delta}- g_{\alpha\delta}g_{\gamma\beta}) \\ \tilde R_{\alpha 0\beta 0} &=& -FF'' g_{\alpha\beta} 
\ea, \qquad
\ba{rcl} 
\tilde R_{\alpha\beta} &=& \left[(N-2)(1-F'^2)-FF''\right]g_{\alpha\beta} \\ \tilde R_{00} &=& -(N-1)F''\!/F .
\ea
\ee}%
Finally we will also need the second covariant derivative of the potential with the covariant derivative being compatible with \eqref{metric}. This is given by
\be \label{nabnabV}
\tilde\nabla^i\tilde\nabla_j V = V''\delta^i_0\delta^0_j + (F'/F)V'(\delta^i_j - \delta^i_0\delta^0_j).
\ee
We proceed by computing different components of the tensor $Q^i_j$ in \eqref{gs}
\be
g^{i l} S_{; l j}= - \delta^i_j\tilde\nabla_\mu\tilde\nabla^\mu-Q^i_j
\ee
whose trace of logarithm appears in the expression for the one-loop effective action.
The general expression for $Q^i_j$ is given by
\be \label{Qon}
Q^i_j = \tilde R^i_{\phantom{i}pjq}\partial_\mu\varphi^p\partial^\mu\varphi^q + \tilde\nabla^i\tilde\nabla_j V  + G^{ip}G_{kq}\delta\Gamma^k_{pj}\tilde\nabla_\mu \partial^\mu\varphi^q + G^{ip}\delta\Gamma^k_{pj}V,_k.
\ee
Because of our choice of coordinates it is more convenient to decompose the indices into 
radial and angular ones. In components we find
\be
Q^0_0 = -FF''g_{\alpha\beta}\partial_\mu\chi^\alpha\partial^\mu\chi^\beta + V''.
\ee
For the pure radial components, which does not get any contribution from the $\delta\Gamma^k_{ij}$ terms in \eqref{Qon}, and
{\setlength\arraycolsep{2pt}
\bea 
Q^\alpha_0
&=& (F''/F+2F'^2/F^2-2F'/hF)\partial_\mu h\partial^\mu\chi^\alpha + (F'/F-/h)\nabla^2_g\chi^\alpha \\
Q^0_\alpha &=& (FF''+2F'^2-2FF'/h)g_{\alpha\beta}\partial_\mu h\partial^\mu\chi^\beta + F(F'-F/h)g_{\alpha\beta}\nabla^2_g\chi^\beta ,
\eea}%
for the mixed components, where the second terms are the $\delta\Gamma^k_{ij}$ contributions required to have a single total field dependence. We also find for the angular components the following expression
{\setlength\arraycolsep{2pt}
\bea 
Q^\alpha_\beta &=& -(F''/F)\delta^\alpha_\beta\partial_\mu h\partial^\mu h + (1-F'^2)(\delta^\alpha_\beta g_{\rho\delta}-\delta^\alpha_\delta g_{\rho\beta})\partial_\mu\chi^\rho\partial^\mu\chi^\delta + (h/F^2)V'\delta^\alpha_\beta \nn\\
&& - ((FF'-h)/F^2)\delta^\alpha_\beta\partial^2 h + FF'((FF'-h)/F^2)\delta^\alpha_\beta g_{\rho\sigma}\partial_\mu\chi^\rho\partial^\mu\chi^\sigma ,
\eea}%
in which the second line is the $\delta\Gamma^k_{ij}$ contribution.
Note that if instead we expand the total field with the connection $\tilde\Gamma^k_{ij}$ compatible with $G_{ij}$, there will be no $\delta\Gamma^k_{ij}$ contribution and we would get
\be 
\tilde Q^i_j = \tilde R^i_{\phantom{i}pjq}\partial_\mu\varphi^p\partial^\mu\varphi^q + \tilde\nabla^i\tilde\nabla_j V, 
\ee
or in components:
{\setlength\arraycolsep{2pt}
\bea
\tilde Q^0_0 &=& -FF''g_{\alpha\beta}\partial_\mu\chi^\alpha\partial^\mu\chi^\beta + V'', \\ 
\tilde Q^\alpha_0 &=& (F''/F)\partial_\mu h\partial^\mu\chi^\alpha, \\ \tilde Q^0_\alpha &=&  FF''g_{\alpha\beta}\partial_\mu h\partial^\mu\chi^\beta,  \\ \tilde Q^\alpha_\beta &=& -(F''/F)\delta^\alpha_\beta\partial_\mu h\partial^\mu h + (1-F'^2)(\delta^\alpha_\beta g_{\rho\delta}-\delta^\alpha_\delta g_{\rho\beta})\partial_\mu\chi^\rho\partial^\mu\chi^\delta + (F'/F)V'\delta^\alpha_\beta.
\eea}%
We emphasize that both $\tilde Q^i_j$ and $Q^i_j$ are covariant and $O(N)$ invariant, but using $\tilde Q^i_j$ to evaluate the one-loop effective action leads to a result which is essentially valid only at the background level. In other words, the extra $\delta\Gamma^k_{ij}$ contributions in \eqref{Qon} make the result valid at all orders of the fluctuation field.
In the following subsections, using dimensional regularization, we will compute explicitly the divergent (pole) terms in the one-loop effective action in $d=2$, where the theory is renormalizable, and in $d=4$ where it is not.
This is the approach normally used in a perturbative study of effective field theories with a mass independent subtraction scheme.

\subsection{$O(N)$ EFT in two space-time dimensions}

To calculate the poles around two space-time dimensions in the one-loop effective action, we take the dimension of space-time to be $d=2-2\epsilon$. Using heat kernel methods, the pole term is given as 
\be \label{div.d=2}
\mathrm{div} \,\frac{i}{2}\mathrm{Tr}\log[\tilde\nabla_\mu\tilde\nabla^\mu + Q] =\frac{1}{2\epsilon(4\pi)}\int_x \mathrm{tr}\,a_1 = - \frac{1}{2\epsilon(4\pi)}\int_x \mathrm{tr}\,Q, \qquad a_1 = -Q.
\ee
The expression is simple in this case and we only need to find the trace of $Q^i_j$. This is given by
\be 
Q^i_i = \tilde R_{pq}\partial_\mu\varphi^p\partial^\mu\varphi^q + \tilde\nabla^i\tilde\nabla_i V + G^{ip}G_{kq}\delta\Gamma^k_{pi}\tilde\nabla_\mu \partial^\mu\varphi^q + G^{ip}\delta\Gamma^k_{pi}V,_k.
\ee
The first term reproduces the well known results \cite{Friedan:1980jf,Friedan:1980jm}. The third term in this expression includes second space-time derivatives of the field. This operator is not of the type that appears in the original action. However, since $Q^i_i$ appears as an integrand, one can integrate by parts to bring this into the form of the kinetic term. This is expected from the renormalizability of the non-linear sigma model in two space-time dimensions. After integration by parts this becomes
\be 
Q^i_i \rightarrow  \tilde R_{pq}\partial_\mu\varphi^p\partial^\mu\varphi^q + \tilde\nabla^i\tilde\nabla_i V - G^{il}G_{kq}\tilde\nabla_p\delta\Gamma^k_{il} \,\partial_\mu\varphi^p\partial^\mu \varphi^q  + G^{\alpha\beta}\delta\Gamma^0_{\alpha\beta}V' .
\ee
Using \eqref{DelGam}, \eqref{Riemann} and \eqref{nabnabV}, this can be evaluated explicitly
{\setlength\arraycolsep{2pt}
\bea 
Q^i_i &=& \left[(N-1)(1-hF'/F)-(1-F'^2) -FF''\right]g_{\alpha\beta}\partial_\mu\chi^\alpha\partial^\mu\chi^\beta  \nn\\
&-& (N-1)(1+F'^2-2hF'/F)/F^2\partial_\mu h\partial^\mu h  + V'' +(N-1) hV'/F^2.
\eea}%
Without the $\delta\Gamma^k_{ij}$ contributions this reduces to
\be
\tilde Q^i_i = \left[(N-2)(1-F'^2)-FF''\right]g_{\alpha\beta}\partial_\mu\chi^\alpha\partial^\mu\chi^\beta -(N-1)(F''\!/F) \partial_\mu h\partial^\mu h + V'' + (N-1)(F'/F)V' .
\ee
Let us consider the particular case of $F(h)$ being a constant independent of $h$ where the geometry of the target space is the cylinder $\mathbb R\times S^{N-1}$.
In such a case (remember that in $d=2$ both $h$ and $F$ are dimensionless) the trace of $Q^i_j$ simplifies to
\be
Q^i_i =(N-2) g_{\alpha\beta}\partial_\mu\chi^\alpha\partial^\mu\chi^\beta  - \frac{(N-1)}{F^2}\partial_\mu h\partial^\mu h  + V'' +(N-1)\frac{h}{F^2} V'.
\ee%
From this result one can observe that in addition to the $O(N)$ symmetry that was expected, at least at the one-loop level, in the kinetic term cylindrical symmetry is also preserved.  
One can verify that the one loop beta function for the coupling $\lambda=1/F$ is given by the well known relation~\cite{polyakov}
\be
\mu \frac{d}{d \mu} \lambda^2=-\frac{(N-2)}{2\pi} \lambda^4.
\ee
where $\mu$ is the additional mass scale required in dimensional regularization.

\subsection{$O(N)$ EFT in four space-time dimensions}

The pole term in four space-time dimensions is more involved. Taking $d=4-2\epsilon$, this is given by
\be \label{oneloopdiv4bis}
\mathrm{div} \,\frac{i}{2}\mathrm{Tr}\log[\tilde\nabla_\mu\tilde\nabla^\mu + Q] = \frac{1}{4\epsilon(4\pi)^{2}}\int_x \mathrm{tr}\,a_2 = \frac{1}{4\epsilon(4\pi)^{2}}\int_x \mathrm{tr}\left[Q^2 +
 \frac{1}{6}\,\tilde R_{\mu\nu\phantom{a}k}^{\phantom{\mu\nu}i} \tilde R^{\mu\nu\, k}_{\phantom{\mu\nu i}j} \right]
\ee
where $a_2$ was given in Eq.~(\ref{a2}) of the previous Section. Here we need to compute the trace of $Q^i_kQ^k_j$. This is given by the following expression
{\setlength\arraycolsep{2pt}
\bea
Q^i_j Q^j_i &=& \tilde R^i_{pjq}\tilde R^j_{kil}\partial_\mu\varphi^p\partial^\mu\varphi^q \partial_\nu\varphi^k\partial^\nu\varphi^l + 2\tilde\nabla^i\tilde\nabla_j V \tilde R^j_{kil}\partial_\mu\varphi^k\partial^\mu\varphi^l + \tilde\nabla^i\tilde\nabla_j V \tilde\nabla^j\tilde\nabla_i V \\
&+& G^{ip}G_{kq}\delta\Gamma^k_{pj}\tilde\nabla_\mu \partial^\mu\varphi^q  
\tilde R^j_{ris}\partial_\nu\varphi^r\partial^\nu\varphi^s + G^{ip}\delta\Gamma^k_{pj}V,_k \tilde R^j_{ris}\partial_\nu\varphi^r\partial^\nu\varphi^s \nn\\
&+& G^{ip}G_{kq}\delta\Gamma^k_{pj}\tilde\nabla_\mu \partial^\mu\varphi^q 
\,\tilde\nabla^j\tilde\nabla_i V + G^{ip}\delta\Gamma^k_{pj}V,_k \tilde\nabla^j\tilde\nabla_i V
+ G^{ip}G^{jr}\delta\Gamma^k_{pj}\delta\Gamma^l_{ri}V,_k V,_l \nn\\
&+& 
G^{ip}G^{jr}G_{kq}G_{ls}\,\delta\Gamma^k_{pj}\delta\Gamma^l_{ri}\tilde\nabla_\mu \partial^\mu\varphi^q\tilde\nabla_\nu \partial^\nu\varphi^s
+2 G^{ip}G^{jr}G_{kq}\delta\Gamma^k_{pj}\delta\Gamma^l_{ri}\tilde\nabla_\mu \partial^\mu\varphi^q V,_l\nn
\eea}%
The first line is the result of exponential expansion with $\tilde\Gamma^k_{ij}$, and the rest are contributions proportional to $\delta\Gamma^k_{ij}$. In components this is
{\setlength\arraycolsep{2pt}
\bea  \label{Q2on}
Q^i_j Q^j_i &=& (FF'')^2 g_{\alpha\beta}g_{\gamma\delta}\partial_\mu\chi^\alpha\partial^\mu\chi^\beta \partial_\nu\chi^\gamma\partial^\nu\chi^\delta - 2 V'' FF''g_{\alpha\beta}\partial_\mu\chi^\alpha\partial^\mu\chi^\beta + (V'')^2 \nn\\
&+& (F''+2(F'/F)(F'-F/h))^2\partial^\mu h\partial_\mu h\, g_{\alpha\beta}\partial^\mu\chi^\alpha\partial^\mu\chi^\beta + (F'-F/h)^2g_{\alpha\beta}\nabla^2_g\chi^\alpha\nabla^2_g\chi^\beta \nn\\
&+& 2(F''+2(F'/F)(F'-F/h))(F'-F/h)\,g_{\alpha\beta}\partial_\mu h\,\partial^\mu\chi^\alpha\,\nabla^2_g\chi^\beta \nn\\
&+& [(1-hF'/F) g_{\rho\delta}\partial_\mu\chi^\rho\partial^\mu\chi^\delta + (h/F^2)V' -(F''/F)\partial_\mu h\partial^\mu h - ((FF'-h)/F^2)\partial^2 h]^2(N-1) \nn\\
&+&  [(1-hF'/F) g_{\rho\delta}\partial_\mu\chi^\rho\partial^\mu\chi^\delta + (h/F^2)V' -(F''/F)\partial_\mu h\partial^\mu h - ((FF'-h)/F^2)\partial^2 h]\times \nn\\
&\times&2(1-F'^2)g_{\rho\delta}\partial_\mu\chi^\rho\partial^\mu\chi^\delta
+ (1-F'^2)^2 g_{\rho\gamma}g_{\sigma\delta}\partial_\mu\chi^\rho\partial^\mu\chi^\delta \partial_\nu\chi^\sigma\partial^\nu\chi^\gamma.
\eea}%
Without the $\delta\Gamma^k_{ij}$ corrections this simplifies to
{\setlength\arraycolsep{2pt}
\bea \label{Qt2on}
\tilde Q^i_j \tilde Q^j_i &=& (FF'')^2 g_{\alpha\beta}g_{\gamma\delta}\partial_\mu\chi^\alpha\partial^\mu\chi^\beta \partial_\nu\chi^\gamma\partial^\nu\chi^\delta - 2 V'' FF''g_{\alpha\beta}\partial_\mu\chi^\alpha\partial^\mu\chi^\beta + (V'')^2 \nn\\
&+& (F'')^2 g_{\alpha\beta}\partial^\mu\chi^\alpha\partial_\mu\chi^\beta \partial^\mu h\partial_\mu h + (1-F'^2)^2 g_{\rho\gamma}g_{\sigma\delta}\partial_\mu\chi^\rho\partial^\mu\chi^\delta \partial_\mu\chi^\sigma\partial^\mu\chi^\gamma \nn\\
&+& [(1-F'^2) g_{\rho\delta}\partial_\mu\chi^\rho\partial^\mu\chi^\delta + (F'/F)V'-(F''/F)\partial_\mu h\partial^\mu h]^2(N-1) \nn\\
&-& 2(1-F'^2)[(1-F'^2) g_{\rho\delta}\partial_\mu\chi^\rho\partial^\mu\chi^\delta + (F'/F)V'-(F''/F)\partial_\mu h\partial^\mu h]g_{\sigma\gamma}\partial_\mu\chi^\sigma\partial^\mu\chi^\gamma.
\eea}%
In order to evaluate \eqref{oneloopdiv4} we also need the trace of the Riemann tensor squared. This can be computed using \eqref{Riemann}
\be  \label{R2on}
\tilde R_{\mu\nu\phantom{a}j}^{\phantom{\mu\nu}i} \tilde R^{\mu\nu\, j}_{\phantom{\mu\nu j}i} = -4(F'')^2g_{\alpha\beta}\,\partial_{[\mu}\chi^\alpha\partial_{\nu]}h\, \partial_{[\mu}\chi^\beta\partial_{\nu]}h + 4(1-F'^2)^2 \partial_{[\mu}\chi^\alpha\partial_{\nu]}\chi^\rho \partial^{[\mu}\chi^\beta\partial^{\nu]}\chi^\sigma g_{\sigma\alpha}g_{\rho\beta} \,.
\ee
The particular case of $N=4$ gives the one-loop divergences in the effective field theory of the Higgs scalar with custodial symmetry. Inserting \eqref{Q2on} and \eqref{R2on} in \eqref{oneloopdiv4} gives a result which allows us to compute not only the one-loop pole term at the background level $\Gamma^{\mathrm{1-loop}}[\varphi,0]$ but also all the fluctuation derivatives $\Gamma^{\mathrm{1-loop}}_{;i_1\cdots i_n}[\varphi,0]$.
Let us mention that contrary to the two dimensional case we see that, even setting $F(h)$ constant in the bare action, the cylindric symmetry is broken at one loop, while $O(N)$ is preserved as expected.

\section{Conclusions}

We have addressed the question if it is possible to use a background-field method in order to construct an off-shell effective action which is invariant
under field reparametrizations in the ultraviolet, and at the same time avoid its separate (or complicated) dependence on the background and fluctuation field,
which is a source of limitation in applying the covariant method beyond the background level. 
Our answer is in the affirmative, provided one uses an exponential splitting generated by a flat connection.
Adopting such splittings which are constructed by geodesics based on a flat connection,
any functional of the total field will solve the splitting Ward identity.
This total field splits into a background and average quantum field in a dynamical independent way. As mentioned at the beginning, the analysis presented here is restricted to scalar theories. 

Another interesting theoretical question regards the preservation of symmetries of a quantum field theory in the infrared. 
Given a symmetry group and its realization on the degrees of freedom, we have explained in Section~\ref{sea} the criteria under which the symmetry of the ultraviolet action is inherited by the effective action. 
Making use of the lemma of Coleman, Wess and Zumino on the linearizability of symmetry realizations, the preservation of the ultraviolet symmetries 
in the infrared is subject to the existence of a fixed point of the group action, either in the field space itself or in an extension of it.
Examples not enjoying such criteria may be embedded in higher dimensional theories. 
Therefore, choosing the extra degrees of freedom to be decoupled from the rest, one can extract some universal features of such theories 
by studying higher dimensional (target space) ones with a covariant approach which also allows for the preservation of ultraviolet symmetries in the infrared. 
An example of this is the $O(N)$ non-linear $\sigma$-model on $S^{N\!-\!1}$ which can be embedded in the space $\mathbb{R}\times S^{N\!-\!1}$ with cylindrical geometry.

We have applied these ideas to a few examples, in particular focusing on the structure of the one loop divergences of the effective action.
Apart from two simple cases with flat target space, we have considered the most general case of a nonlinear $\sigma$-models. 
We have given some general results in this context, including the expression for the one-loop divergences and furthermore we have shown explicitly, 
that the one-loop effective potential is single-field dependent, i.e. even a background computation leads, simply by replacing the background field by the total field, 
to the construction of a covariant (off-shell) effective action with the correct dependence on the fluctuating fields to all orders.

Finally, we have discussed in some details $O(N)$ invariant effective field theories of $N$ scalars. This provides an example of a theory with curved target space for which the standard Vilkovisky-DeWitt approach leads to a background-dependent effective action. It also fulfills the criterion discussed in Section \eqref{sea} so that $O(N)$ invariance is present in the effective action as well. This is of interest for the analysis of the Higgs sector of the standard model and its extensions.
In particular it applies to the so called Higgs Effective Field Theory (HEFT).
We plan to study in future analyses such effective field theories, also with the full electroweak symmetry, as well as the case of supersymmetric extensions of the SM. 

Regarding the issue of spontaneous symmetry breaking, it is worth emphasizing that the single-field and covariant construction presented in this work concerns the off-shell effective action, and the notion of vacuum is by definition independent of this choice (as are scattering amplitudes).
Just like the linear split, one is free, for instance, to choose the background field also in the non-linear case to coincide with the vacuum of the theory, in which case $\xi^i=0$ will correspond to a possible symmetry-breaking point.

We have not addressed the extension to the more subtle case of gauge theories here. Even though (at least) Yang-Mills theory possesses a flat geometry, maintaining a local description has been the main obstacle in dealing with gauge theories. In fact, in this case the approach presented here can coincide with that of Vilkovisky-DeWitt. We will leave for a future investigation the exploration of possible other ideas in this direction. 

Throughout this work we have chosen dimensional regularization along with $\overline{\mathrm{MS}}$. However, as briefly pointed out in Section \eqref{ssec.1f}, the main features of our approach, i.e. covariance and single-field dependence, are independent of the scheme of perturbative regularization. Further, we believe that our approach may prove useful also for non perturbative analysis within the framework of the Wilsonian functional renormalization group.
One of the advantages of having a covariant and single-field dependent description is the strong constraint on the possible operators appearing in the effective action. Investigation along this direction is presented elsewhere \cite{SV}.


\appendix

\section{Exponential splitting with a flat connection}

In this section we explain how the splitting introduced in section \eqref{fqbs} is related to the splitting with the exponential map.
Specifically we show that for a flat torsion-free connection $\Gamma^k_{ij}$ the exponential map, which includes in its expansion all powers of the fluctuation field,
can be written as a function of a simple combination of the background and fluctuation field which is linear in the fluctuations.
More explicitly, denoting by $[Exp_\varphi\, \xi]_\Gamma$ the exponential map based on the connection $\Gamma^k_{ij}$ we will show that 
\be \label{fexp}
[Exp_\varphi\, \xi]_\Gamma^i \equiv \varphi^{i}+\xi^{i} -\sum_{n=2}^{\infty}\frac{1}{n!}\, \Gamma^{i}_{i_1 i_2 \ldots i_n}(\varphi)\,\xi^{i_1}\cdots\xi^{i_n} = \left[f^{-1}\left(f(\varphi)+ \partial f\xi\right)\right]^i ,
\ee
where $f$ is a coordinate transformation function under which the connection $\Gamma^k_{ij}$ vanishes, and $\Gamma^{i}_{i_1 i_2 \ldots i_n}$ are the covariant derivatives of the connection ignoring the upper index $i$. 
To show this, we need to express the connection in terms of the function $f$. The flat connection is the one generated from the vanishing connection by the transformation $x^i\rightarrow x'^i$ 
\be 
x'^i = f^{-1}(x), \qquad  \bar U^i_a = \frac{\partial x'^i}{\partial x^a}.
\ee
This is found in the following way
\be 
0 \rightarrow
\bar U^k_b\,(\bar U^{-1})^a_i\,\partial_a (\bar U^{-1})^b_j = \bar U^k_b\,\partial'_i (\bar U^{-1})^b_j = \frac{\partial x'^k}{\partial x^b} \frac{\partial x^b}{\partial x'^i\partial x'^j} = (f^{-1})^k\!\!,_b f^b\!\!,_{ij} \equiv \Gamma^k_{ij}(x).
\ee
An expansion in the fluctuation field of the function on the right hand side of Eq.~\eqref{fexp} gives
{\setlength\arraycolsep{2pt}
\bea 
\left[f^{-1}\left(f(\varphi)+ \partial f\xi\right)\right]^i &=&  \sum_{n=0}^{\infty}\frac{1}{n!}\left.(f^{-1})^i\!,_{i_1\cdots i_n}\right|_{f(\varphi)}\, f^{i_1}\!\!\!,_{j_1}\cdots f^{i_n}\!\!\!,_{j_n}\, \xi^{j_1} \cdots \xi^{j_n},
\eea}%
where $f^{i_k}\!\!\!,_{j_k}$ are evaluated at the background. It is easy to see that the first two terms in this expansion match those in \eqref{fexp}
\be \label{ftt}
(f^{-1})^i(f(\varphi)) = \varphi^i, \qquad
(f^{-1}),_k^i(f(\varphi))\,f^k\!\!\!,_j(\varphi) = \delta^i_j, \qquad
\ee
To complete the proof of \eqref{fexp} we need to show that for $n\geq 2$
\be \label{induction}
(f^{-1})^i\!,_{i_1\cdots i_n}\left(f(\varphi)\right)\, f^{i_1}\!\!\!,_{j_1}\cdots f^{i_n}\!\!\!,_{j_n} = -\Gamma^{i}_{i_1 i_2 \ldots i_n}(\varphi).
\ee
This is easily shown by induction. Taking another derivative of the right equation in \eqref{ftt} we find
\be 
0= (f^{-1}),_{mn}^i\,f^m\!\!\!,_p\,f^n\!\!\!,_q + (f^{-1}),_k^i\,f^k\!\!\!,_{pq}= (f^{-1}),_{mn}^i\,f^m\!\!\!,_p\,f^n\!\!\!,_q + \Gamma^i_{pq}
\ee
which gives \eqref{induction} for $n=2$. This can also be rewritten as
\be \label{formula}
f^b\!\!,_{ij} = f^b\!\!,_k\Gamma^k_{ij}(x) \qquad (\nabla_i\nabla_j f^b=0),
\ee
where in the equation in the parenthesis $b$ is not considered as a tensor index. Assuming that \eqref{induction} holds for some $n>2$
\be \label{ind1}
0=(f^{-1})^i\!,_{i_1\cdots i_n}\left(f(\varphi)\right)\, f^{i_1}\!\!\!,_{j_1}\cdots f^{i_n}\!\!\!,_{j_n} + \Gamma^i_{j_i\cdots\j_n} 
\ee
we need to show that it also holds for $n+1$. Simply taking another derivative of the above equation we get
{\setlength\arraycolsep{2pt}
\bea 
0 &=& (f^{-1})^i\!,_{i_1\cdots i_{n+1}}\, f^{i_1}\!\!\!,_{j_1}\cdots f^{i_{n+1}}\!\!\!,_{j_{n+1}} \\
&+& (f^{-1})^i\!,_{i_1\cdots i_n}\, f^{i_1}\!\!\!,_{j_1j_{n+1}}\cdots f^{i_n}\!\!\!,_{j_n} + \cdots +  (f^{-1})^i\!,_{i_1\cdots i_n}\left(f(\varphi)\right)\, f^{i_1}\!\!\!,_{j_1}\cdots f^{i_n}\!\!\!,_{j_nj_{n+1}} + \partial_{j_{n+1}}\Gamma^i_{j_i\cdots j_n} \nn
\eea}
The second line can be rewritten as
{\setlength\arraycolsep{2pt}
\bea 
&& \partial_{j_{n+1}}\Gamma^i_{j_i\cdots j_n} + (f^{-1})^i\!,_{i_1\cdots i_n}\, f^{i_1}\!\!\!,_k\Gamma
^k_{j_1j_{n+1}}\cdots f^{i_n}\!\!\!,_{j_n} + \cdots +  (f^{-1})^i\!,_{i_1\cdots i_n}\left(f(\varphi)\right)\, f^{i_1}\!\!\!,_{j_1}\cdots f^{i_n}\!\!\!,_k\Gamma
^k_{j_nj_{n+1}} \nn\\[2mm]
&=& \partial_{j_{n+1}}\Gamma^i_{j_i\cdots j_n} - \Gamma^i_{k j_2\cdots j_n}\Gamma
^k_{j_1j_{n+1}} - \cdots -  \Gamma^i_{j_1\cdots j_n k}\Gamma
^k_{j_nj_{n+1}} = \nabla_{j_{n+1}}\Gamma^i_{j_i\cdots j_n} \equiv  \Gamma^i_{j_i\cdots j_n j_{n+1}},
\eea}%
where we have used \eqref{formula} in the first line and \eqref{ind1} in the second line. This completes the proof of \eqref{fexp}.
Given the above results, the more general split \eqref{split1} has also an interpretation in terms of the exponential map but with a redefined vector $\tilde\xi = (\partial f)^{-1}U\xi$
\be  
f^{-1}\left(f(\varphi)+ U\xi\right) = f^{-1}(f(\varphi)+ \partial f\tilde \xi) = [Exp_\varphi\, \tilde\xi]_\Gamma.
\ee
In terms of $\tilde \xi$ the differential equation \eqref{linear split} takes the form
\be
\tilde\xi^k \nabla_k \tilde \xi^i+\tilde \xi^i=0.
\label{geodesiceq}
\ee
where $\nabla$ is the covariant derivative related to the flat connection $\Gamma^k_{ij}$. We shall mostly consider the case $U^i_a(\varphi)=[(\partial f(\varphi))^{-1}]^i_a$
for which $\tilde \xi=\xi$ has the specific dependence on the total ($\phi$) and the background ($\varphi$) fields given in Eq.~(\ref{solgeoeq}).
Finally notice that defining as usual the biscalar quantity $\sigma=\frac{1}{2} ({\rm geodesic \,\, distance})^2$, and recalling the form of the induced metric $g_{ij}=f^a_{,i}f^a_{,j}$,
one trivially has $\sigma(\phi,\varphi) =\frac{1}{2} \left( f^a(\phi)\!-\!f^a(\varphi)\right)^2$.
This also follows starting from the result in Euclidean space and changing parametrization from the case $f={\rm id}$.
One can check also the well known expression for the tangent vector as
\be
\xi^i = g^{ik} \sigma_{,k}.
\ee
\section{Single-field dependence: an explicit check}
Let us take the action \eqref{nlsm}
and compute the one-loop divergences of the effective action at next to leading (first) order in the fluctuation field.
For this purpose we need to expand the action up to third order in the fluctuations. This can be done, using \eqref{glspexp} in the potential, and making the replacements
{\setlength\arraycolsep{2pt}
\bea
g_{ij}(\phi) &\rightarrow & g_{ij}(\varphi)+ \xi^p\nabla\!_p\, g_{ij}+\cdots \nn\\
\partial_\mu\phi^i &\rightarrow & \partial_\mu\varphi^i+\nabla_\mu\xi^i +\cdots 
\eea}%
in the kinetic term. For simplicity we focus on quantum correction to the potential only. Keeping terms which contribute to the potential at one-loop 
(keeping terms with no space-time derivatives of the background field and at most two space-time derivatives of the fluctuations) and up to third order in the fluctuation fields we find
\be
{\textstyle{\frac{1}{2}}}\,g_{ij}(\phi)\,\partial_\mu\phi^i\partial^\mu\phi^j =
{\textstyle{\frac{1}{2}}}\,\left(g_{ij}(\varphi)+ \xi^p\nabla\!_p\, g_{ij}\right)\,\partial_\mu\xi^i\,\partial^\mu\xi^j + \cdots
\ee
\be
V(\phi) =
V(\varphi) + \xi^p\nabla\!_p\,V(\varphi) + {\textstyle{\frac{1}{2}}}\xi^p\xi^q\nabla\!_p\nabla\!_q\,V(\varphi) + {\textstyle{\frac{1}{3!}}}\xi^p\xi^q\xi^r\nabla\!_p\nabla\!_q\nabla\!_r\,V(\varphi) + \cdots
\ee
Taking the second fluctuation derivative of the action and setting the fields to constants we get
{\setlength\arraycolsep{2pt}
\bea \label{sij}
S_{;ij}[\phi] &=& -\left(g_{ij}(\varphi)+ \xi^p\nabla\!_p\, g_{ij}\right)\square - \nabla\!_{(i}\nabla\!_{j)}\,V(\varphi) - \xi^r\nabla\!_{(i}\nabla\!_j\nabla\!_{r)}\,V(\varphi) + \cdots \nn\\
&=& \left(g_{ik}(\varphi)+ \xi^p\nabla\!_p\, g_{ik}\right)\left[-\square \delta^k_j - g^{kq}(\varphi)\nabla\!_{(q}\nabla\!_{j)}\,V(\varphi) \right. \nn\\
&& \left. \phantom{\left(g_{ik}(\varphi)+ \xi^p\nabla\!_p\, g_{ik}\right)} -\xi^p\nabla\!_p\,g^{kq}(\varphi)\nabla\!_{(q}\nabla\!_{j)}\,V(\varphi) - \xi^r g^{kq} \nabla\!_{(q}\nabla\!_j\nabla\!_{r)}\,V(\varphi) + \cdots \right]
\eea}%
where we have used the fact that to first order in the fluctuations the inverse of the quantity behind the brackets in \eqref{sij} is 
\be 
g^{kq}(\varphi) + \xi^p\nabla\!_p\, g^{kq}.
\ee
The divergent part of the one-loop effective action is proportional to the trace of the terms in front of $\square$. Up to first order in the fluctuations this is
\be \label{fo}
g^{kq}(\varphi)\nabla\!_{(q}\nabla\!_{j)}\,V(\varphi)\left[g^{jp}(\varphi)\nabla\!_{(p}\nabla\!_{k)}\,V(\varphi) +  2 \xi^p\nabla\!_p\,g^{jq}(\varphi)\nabla\!_{(q}\nabla\!_{k)}\,V(\varphi)\,V(\varphi) + 2 \xi^r g^{jq} \nabla\!_{(q}\nabla\!_k\nabla\!_{r)}\,V(\varphi)\right].
\ee
At the background this becomes
\be \label{fob}
g^{kq}(\varphi)\nabla\!_{(q}\nabla\!_{j)}V(\varphi)\; g^{jp}(\varphi)\nabla\!_{(p}\nabla\!_{k)}V(\varphi).
\ee
Now, we would like to see if replacing $\varphi^i\rightarrow \phi^i = \varphi^i + \xi^i + \cdots $ in the background result \eqref{fob} and expanding up to first order in the fluctuations we can reconstruct the first order result \eqref{fo}. The result of the expansion of \eqref{fob} up to first order is
\bea
\!\!\!g^{kq}(\varphi)\nabla\!_{(q}\nabla\!_{j)}V(\varphi)\left[g^{jp}(\varphi)\nabla\!_{(p}\nabla\!_{k)}V(\varphi) + 2\xi^s\nabla_s g^{jp}(\varphi)\,\nabla\!_{(p}\nabla\!_{k)}V(\varphi) + 2\xi^s g^{jp}(\varphi)\nabla_s \nabla\!_{(p}\nabla\!_{k)}V(\varphi)\right].
\label{forec}
\eea
The first two terms here are the same as the first two terms in \eqref{fo}. 
Therefore the two expression \eqref{fo} and \eqref{forec} match if 
\be 
\nabla_r \nabla\!_{(q}\nabla\!_{k)}V(\varphi) = \nabla\!_{(q}\nabla\!_k\nabla\!_{r)}\,V(\varphi).
\ee
This is not generally the case, but it is true if the connection is flat, so that the covariant derivatives commute with each other.\\
%


\end{document}